\documentclass[aps,prb,superscriptaddress,twocolumn,showpacs,tightenlines,
amsmath,amsfonts] {revtex4}

\usepackage{graphics,bm}

\newcommand{\sump}{\mathop{{\sum}'}}
\newcommand{\prodp}{\mathop{{\prod}'}}
\newcommand{\prodpp}{\mathop{{\prod}''}}

\begin{document}

\author{O. A. Tretiakov}
\affiliation{Department of Physics, Duke University, Durham, NC 27708}

\author{K. A. Matveev}
\affiliation{Materials Science Division, Argonne National Laboratory,
Argonne, IL 60439}
\affiliation{Department of Physics, Duke University, Durham, NC 27708}

\date{November 2, 2004}

\title{Stochastic current switching in bistable resonant tunneling
  systems}

\begin{abstract}
  Current-voltage characteristics of resonant-tunneling structures often
  exhibit intrinsic bistabilities.  In the bistable region of the $I$-$V$
  curve one of the two current states is metastable.  The system switches
  from the metastable state to the stable one at a random moment in time.
  The mean switching time $\tau$ depends exponentially on the bias
  measured from the boundary of the bistable region $V_{th}$.  We find
  full expressions for $\tau$ (including prefactors) as functions of bias,
  sample geometry, and in-plane conductivity.  Our results take universal
  form upon appropriate renormalization of the threshold voltage $V_{th}$.
  We also show that in large samples the switching initiates inside, at
  the edge, or at a corner of the sample depending on the parameters of
  the system.
\end{abstract}


\maketitle

\hyphenation{pre-factor}

\section{Introduction}

Recent advances in experimental techniques have made possible the study of
fast stochastic processes such as dynamic current switching in resonant
tunneling structures.  The electron transport in these devices has
attracted a lot of attention since the pioneering work of Tsu and
Esaki.~\cite{Tsu} The interest was further stimulated by the discovery of
the phenomenon of intrinsic bistability~\cite{Goldman1, Alves1, Goldman2,
  Hayden:exp, Mendez} in double-barrier resonant tunneling structures
(DBRTS).  Other resonant tunneling structures, such as superlattices, are
also known to show bistable behavior.~\cite{Grahn94, Grahn96, Grahn98,
  Teitsworth} Recent experiments~\cite{Grahn96, Grahn98, Teitsworth}
established that in the bistable region one of the current states is
metastable, and the switching to the stable state was studied.  Both the
mean switching time and its distribution function were
measured.~\cite{Teitsworth}

\begin{figure}[b]
  \resizebox{.35\textwidth}{!}{\includegraphics{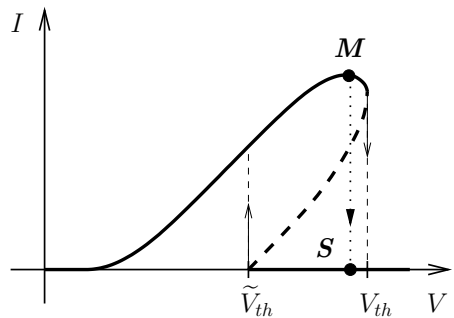}}
\caption{\label{fig:IVcurve} The $I$-$V$ curve of the DBRTS.  The bistable
  region is present in the range of bias between ${\widetilde V}_{th}$ and
  $V_{th}$.  The bold dashed line corresponds to the unstable current
  state.}
\end{figure}

The existence of intrinsic bistability is well understood
theoretically.~\cite{Sheard88, numerical89, Anda93, Blanter99} It was
shown~\cite{Blanter99} that in a certain range of bias, ${\widetilde
  V}_{th} < V < V_{th}$, for every value of $V$ the current can take two
different values, see Fig.~\ref{fig:IVcurve}.  If one increases the bias
starting from any value below ${\widetilde V}_{th}$, the current follows
the upper branch of the $I$-$V$ curve shown in Fig.~\ref{fig:IVcurve}
until $V$ reaches $V_{th}$, where the current switches to the lower
branch.  On the other hand, if one decreases the bias from the values
greater than $V_{th}$, the current follows the lower branch and then
switches to the upper branch at ${\widetilde V}_{th}$.

\begin{figure}
  \resizebox{.41\textwidth}{!}{\includegraphics{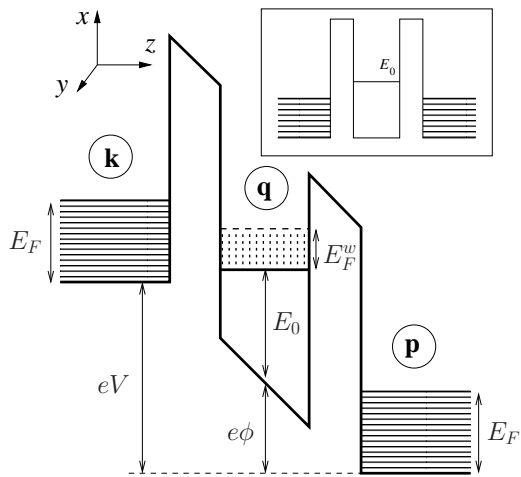}}
\caption{\label{fig:structure} Schematic potential profile of the
  double-barrier resonant tunneling structure.  The structure consists of
  a quantum well separated from two leads by tunneling barriers.  The
  electrons with three-dimensional wavevectors $\mathbf{k}$ and
  $\mathbf{p}$ fill all the states up to the Fermi energies $E_F$ in the
  left and right leads, respectively.  In the quantum well the motion of
  electrons in the $z$-direction is quantized, and the electrons with
  two-dimensional transverse wavevectors $\mathbf{q}$ occupy all the
  states up to the Fermi energy $E_F^w$.  The inset shows the potential
  profile at zero bias.}
\end{figure}

The bistability can be understood by considering the potential profile of
the DBRTS schematically shown in Fig.~\ref{fig:structure}.  If the level
$E_0$ in the quantum well is below the bottom of the conduction band of
the left lead, tunneling into the well is not possible, and the current
through the heterostructure is zero. In this case the charge in the well
$Q=0$.  However, if a non-zero charge $Q$ is added to the well, the level
$E_0$ rises due to the charging effects and may become higher than the
bottom of the conduction band of the left lead.  Then, another steady
state of current is possible.  In this state the current into the well
from the left lead is compensated by the current out of the well through
the right barrier.  Thus, it is possible to have two different current
states at the same bias.  (See, e.g., points \textbf{\textit{M}} and
\textbf{\textit{S}} on the $I$-$V$ curve, Fig.~\ref{fig:IVcurve}.)

The electric current in the device fluctuates, because the electrons
tunnel in and out of the well at random moments in time.  The resulting
shot noise of current through the heterostructure gives rise to the
metastability of some current states.  The two solid lines in
Fig.~\ref{fig:IVcurve} correspond to the most probable values of current
at a given bias.  These two branches are stable, i.e., any \textit{small}
fluctuation of current near a solid line will decay with time, and the
current will return to its value at the solid line.  The dashed line
between ${\widetilde V}_{th}$ and $V_{th}$ corresponds to the unstable
state.  Here any deviation from the dashed line which raises or lowers the
current will switch the system to the upper or lower stable current state,
respectively.

Qualitative understanding of the metastability can be achieved by
considering the system at a bias near $V_{th}$, e.g., point
\textbf{\textit{M}} on the upper branch of the $I$-$V$ curve,
Fig.~\ref{fig:IVcurve}.  Then, as one can see from Fig.~\ref{fig:IVcurve},
a relatively small fluctuation can shift the current below the dashed line
corresponding to the unstable state.  If that happens, the system switches
to the lower branch.  The opposite process is much less probable, since
the distance from the lower branch to the dashed line is much larger than
that from the upper branch.  Therefore, the lower branch is stable, and
the system remains in that state.

The dependence of the mean switching time $\tau$ on the bias was addressed
theoretically in Ref.~\onlinecite{paper1}.  It was shown that near the
threshold voltage $V_{th}$ the logarithm of $\tau$ behaves as
\begin{equation}
\label{paper1}
\ln\tau \propto
\left\{
\begin{array}{ll}
(V_{th}-V)^{3/2}, &L\ll r_0 ,\\
V_{th}-V, &L\gg r_0 .
\end{array}
\right.
\end{equation}
Here $L$ is the size of the sample, $r_0 \propto \sigma^{1/2}
(V_{th}-V)^{-1/4}$ is a characteristic length scale, and $\sigma$ is the
in-plane conductivity.  In small samples, $L\ll r_0$, the current switches
from the metastable state to the stable one simultaneously over the entire
area of the device.  On the other hand, in large samples, $L\gg r_0$, the
switching is initiated in a small critical region of radius $r_0$.  After
the switching has occurred in that region, it extends rapidly to the rest
of the sample.

In this paper we show that if the sample is large, the switching can
initiate not only inside, but also at the edge of the device.  The latter
process tends to be more efficient, since the exponential in the
respective expression for $\tau$ is smaller than in the case of switching
far from the edges of the sample (Sec.~\ref{sec:large}).  On the other
hand, the switching at the edge can be initiated anywhere along the
boundary of the device, and thus the prefactor of the switching rate
$1/\tau$ due to these processes is proportional to the perimeter $\sim L$.
Similarly, the prefactor of the rate of switching inside the device is
proportional to the area $\sim L^2$, which makes these processes more
efficient in larger samples.

We obtain analytically the full expressions for $\tau$, including the
preexponential factors.  Apart from the dependence on sample dimensions,
the calculation of the prefactors reveals the non-trivial dependence of
the threshold voltage $V_{th}$ on the degree of disorder of the sample.
Formal evaluation of the prefactors in the case of non-uniform electron
density in the well results in ultraviolet divergences.  Similar
divergences appear in quantum field theory, where they are eliminated with
the use of a renormalization procedure.~\cite{ZinnJustin, Coleman1} The
application of a similar technique to our problem leads to the
renormalization of the threshold voltage which depends strongly on the
conductivity of the quantum well
(Secs.~\ref{sec:renormalization},~\ref{sec:large}).  Upon this
renormalization $\ln\tau$ in large samples acquires logarithmic
corrections to its linear voltage dependence.

The paper is organized as follows.  In Sec.~\ref{sec:2} we obtain the
Fokker-Planck equation for tunneling in DBRTS which completely describes
the electron transport in small samples.  This equation enables us to find
a simple result for the mean switching time in these samples.  In
Sec.~\ref{sec:3} we derive the Fokker-Planck equation for the case of
large samples which describes the dynamics of electron density in the well
due to both the diffusion in the plane of the well and tunneling between
the well and the leads.  We use it to investigate the effect of weak
density fluctuations on the decay of metastable current state in small
samples (Sec.~\ref{sec:decay}) and to study the switching in large samples
(Sec.~\ref{sec:large}).  The application of our theory to the existing and
future experiments is discussed in Sec.~\ref{sec:discussion}.

\section {\label{sec:2} Fokker-Planck equation for tunneling in DBRTS}

The bistable current-voltage characteristic of DBRTS was studied
theoretically in Refs.~\onlinecite{Sheard88, numerical89, Anda93,
 Blanter99}.  The $I$-$V$ curve shows the dependence of the
\textit{average} current on voltage applied to the device.  In addition,
shot noise was studied in the regime of small
fluctuations.~\cite{numerical89, Anda93, Blanter99} On the other hand, the
switching between the branches of the $I$-$V$ curve is caused by large
fluctuations of current.  In this section we use the model of
Ref.~\onlinecite{Blanter99} to derive the Fokker-Planck equation for
tunneling in DBRTS, which accounts for these large fluctuations, and thus
describes the switching.

The model is illustrated in Fig.~\ref{fig:structure}.  The well is
extended in $x$-$y$ plane.  The motion in $z$-direction in the well is
quantized, and the well is assumed to have only one resonant level of
energy $E_0$.  The two-dimensional wavevectors in the well are denoted by
$\mathbf{q}$.  The left and right leads are three-dimensional; the
wavevectors of electrons are denoted by $\mathbf{k}$ and $\mathbf{p}$,
respectively.  The conduction bands in the leads are occupied up to the
Fermi energy $E_F$.  In typical devices $E_0$ is of the order of $E_F$;
for definiteness we assume $E_0 > E_F$.  The temperature $T$ is assumed to
be small compared to $E_F$ and $eV$.  The well is separated from the leads
by two tunneling barriers with the transmission coefficients much smaller
than unity.

In Ref.~\onlinecite{Blanter99} the tunneling through the double barrier
was described quantum-mechanically using the Breit-Wigner formula.  The
level widths with respect to the decay to the right and left leads
$\Gamma_{L}$, $\Gamma_{R}$ there were eventually taken to be much smaller
than all other relevant energy scales.  We make this assumption from the
beginning, and describe the electron transport through the barriers using
the sequential tunneling approach.  This method is an alternative to the
use of the Breit-Wigner formula; it enables us to discuss both the $I$-$V$
characteristic and the large fluctuations of current.

In order to have a steady state of non-zero current in the device, the
electrochemical potential in the well should lie between those in the left
and right leads, i.e.,
\begin{equation}
\label{condJ}
eV + E_F > E_0 + e\phi + E_F^w > E_F .
\end{equation}
Here $E_F^w = \hbar^{2}q_F^2 /2m$ is the Fermi energy in the well with $m$
being the effective mass.  Then, in the limit of low temperature the
inequalities~(\ref{condJ}) dictate that the tunneling is possible only in
one direction, namely, from left to right, Fig.~\ref{fig:structure}.  The
probability to tunnel through a barrier is given by the Fermi golden rule.
The rates of electron tunneling into the well $J_L$ and out of the well
$J_R$ take the form:
\begin{eqnarray}
\label{rateL}
J_{L}& = & \frac{4\pi}{\hbar} \sum_{\mathbf{q k}}
|t_{k_{z}}|^{2} \delta_{\mathbf{q k}_{\parallel}}
f_{\mathbf{k}} (1-f_{\mathbf{q}})
\nonumber \\
&&\times \delta (eV + E(k_{z}) - E_0 - e\phi ),
\\
\label{rateR}
J_{R}& = & \frac{4\pi}{\hbar} \sum_{\mathbf{q p}}
|t_{p_{z}}|^{2} \delta_{\mathbf{q p}_{\parallel}}
 f_{\mathbf{q}} \delta ( E_{0} + e\phi - E(p_{z}) ).
\end{eqnarray}
Here $E(k) = \hbar^{2} k^{2}/2m$; $f_{\mathbf{k}}$ and $f_{\mathbf{q}}$
are the Fermi occupation numbers in the left lead and the quantum well,
respectively.  In Eq.~(\ref{rateR}) we used the fact that the Fermi
occupation numbers in the right lead $f_{\mathbf{p}}=0$ at energies above
$E_{0} + e\phi$, because $E_0 >E_F$.
Expressions~(\ref{rateL}),~(\ref{rateR}) include an additional factor of
$2$, which accounts for electron spins.  The matrix elements $t_{p_{z}}$
($t_{k_{z}}$) describe the transitions between the resonant level in the
well and the state with $z$-component of the wavevector $p_z$ ($k_z$) in
the right (left) lead.  The conservation of the transverse momentum is
taken into account by Kronecker deltas.

To simplify the expression for the tunneling rate~(\ref{rateR}) we use
$\delta_{\mathbf{qp}_{\parallel}}$ to remove the sum over
$\mathbf{p}_{\parallel}$.  The remaining sum over $\mathbf{q}$ of Fermi
function $f_{\mathbf{q}}$ gives exactly the number of electrons in the
well with a given spin $N/2$.  Then Eq.~(\ref{rateR}) reduces to
\begin{equation}
\label{JR}
J_{R} = \frac{\Gamma_{R}}{\hbar} N.
\end{equation}
Here $\Gamma_{R}$ is the level width with respect to tunneling into the
right lead.  We define the level widths for the two possible tunneling
processes as
\begin{subequations}
\begin{eqnarray}
\label{GammaL}
\Gamma_{L}& = &2\pi\sum_{k_{z}} |t_{k_{z}}|^{2}
\delta (eV + E(k_{z}) - E_{0} - e\phi ),\\
\label{GammaR}
\Gamma_{R}& = &2\pi\sum_{p_{z}} |t_{p_{z}}|^{2}
\delta ( E_{0} + e\phi - E(p_{z}) ).
\end{eqnarray}
\end{subequations}

To find $J_L$ we use the Kronecker delta to remove the sum over
$\mathbf{k}_{\parallel}$ in Eq.~(\ref{rateL}), while the value of
$k_{z}^{2} = (2m/ \hbar^{2}) (E_0 +e\phi -eV)$ is fixed by the delta
function.  At $\mathbf{k} = (\mathbf{q}, k_z)$ and $T\to 0$ the sum over
$\mathbf{q}$ of $f_{\mathbf{k}} (1-f_{\mathbf{q}})$ can be easily
evaluated, and gives $(S/4\pi)(k_{F}^{2} - k_{z}^{2} - q_{F}^{2})$ under
the condition (\ref{condJ}), where $S$ is the area of the sample.  Then,
the expression~(\ref{rateL}) can be simplified as follows
\begin{equation}
\label{JL}
J_L  = \frac{\Gamma_{L}}{\hbar} \left[ \frac{Sm}{\pi\hbar^2}
(E_F + eV - E_0 - e\phi ) - N \right].
\end{equation}
Here we used the expression $N = Sq_{F}^{2}/2\pi$ for the total number of
electrons in the well.  Note that at $eV > e\phi +E_0$ the level
width~(\ref{GammaL}) vanishes, and thus $J_L =0$.

In the sequential tunneling approximation the average number of electrons
in the well can be determined from the condition $J_L = J_R$,
\begin{equation}
\label{N}
N = \frac{Sm}{\pi\hbar^2} \frac{\Gamma_{L}}{\Gamma_{L}
+ \Gamma_{R}} (E_{F} + eV - E_{0} -e\phi).
\end{equation}
One cannot directly obtain $N$ from Eq.~(\ref{N}), since the potential
$\phi$ depends on the number of electrons in the well.  Considering the
barriers as two capacitors, one finds from electrostatics the following
expression for the electric potential of the well
(Fig.~\ref{fig:structure}),
\begin{equation}
\label{ephi}
\phi = \frac{V}{2} + \frac{eN}{2C}.
\end{equation}
Here we assumed for simplicity that the capacitances of the left and right
barriers are equal to each other, and denoted the capacitance of each
barrier as $C$.

One can obtain the current-voltage characteristic of the DBRTS by
repeating the following steps of Ref.~\onlinecite{Blanter99}.  First, one
notices that the level widths are energy dependent:
\begin{subequations}
\begin{eqnarray}
\label{GL}
\Gamma_{L} &=& g_{L} \sqrt{E_{0}(E_0 -eV +e\phi)},
\\
\label{GR}
\Gamma_{R} &=& g_{R} \sqrt{E_{0}(E_{0} +e\phi)},
\end{eqnarray}
\end{subequations}
where $g_{L,R}$ are dimensionless constants.  Since $\Gamma_{L}$ and
$\Gamma_{R}$ depend on $\phi$, they are also functions of $N$.  Therefore
to find $N$ one has to solve the pair of equations~(\ref{N})
and~(\ref{ephi}).  The latter leads to an equation on $N$, which has three
solutions in the bistable region. One of the solutions corresponds to the
average number of electrons on the unstable branch, while the other two
correspond to $N$ on the lower ($N=0$) and upper stable branches.  Upon
substitution of $N$ into Eq.~(\ref{JR}) one finds the dependence of the
average current on bias, i.e., the bistable $I$-$V$
curve,~\cite{Blanter99} which is schematically shown in
Fig.~\ref{fig:IVcurve}.

To account for the noise, we go one step further and write the master
equation for the time evolution of the distribution function $P(N,t)$ of
the number of electrons in the well $N$.  In terms of the tunneling
rates~(\ref{rateL}) and~(\ref{rateR}), the master equation for $P(N,t)$
takes the form
\begin{eqnarray}
\label{masterCandD}
\frac{\partial}{\partial t}  P(N,t) &=& P(N-1,t) J_{L}(N-1)
\nonumber \\
&&+ P(N+1,t) J_{R}(N+1)
\nonumber \\
&&- P(N,t) [J_{L}(N) + J_{R}(N)].
\end{eqnarray}
The first two terms in the right-hand side of Eq.~(\ref{masterCandD})
account for the processes which increase the probability to have $N$
electrons in the well, while the last term corresponds to the opposite
processes.

In this section we consider the samples of large in-plane conductivity
where the density in the well is uniform.  Therefore, in the steady state
of non-zero current the total number of particles in the well is
proportional to the area of the sample.  The linear dimensions of the
sample are assumed to be large compared to the Bohr radius in the
semiconductor.  Thus the total number of electrons in the well is large,
$N\gg 1$, and one can expand Eq.~(\ref{masterCandD}) in $1/N$.  Keeping
the terms up to the second order, the master equation
reduces~\cite{Landauer62} to
\begin{equation}
\label{tunnelingFPE}
\frac{\partial}{\partial t} P(N,t)
= -\frac{\partial }{\partial N}[A(N)P(N,t)]
+ \frac{1}{2}\frac{\partial^2}{\partial N^2}[B(N)P(N,t)].
\end{equation}
Here $A(N) = J_{L}(N) - J_{R}(N)$ and $B(N) = J_{L}(N) + J_{R}(N)$.
Equation~(\ref{tunnelingFPE}) is known as the Fokker-Planck equation, and
is widely used for the description of various stochastic processes, see,
e.g., Refs.~\onlinecite{vanKampen, Gardiner}.

The stationary solution of Eq.~(\ref{tunnelingFPE}) can be easily
obtained:
\begin{equation}
\label{potential}
P_{0}(N) = \frac{\mbox{const}}{B(N)}e^{-U(N)},
\quad U(N) = -2\int_0^N \frac{A(N')}{B(N')}\,dN',
\end{equation}
Ref.~\onlinecite{vanKampen}.  The extrema of function $U(N)$ are
determined by the condition $A\equiv J_L-J_R =0$, which we used above to
find the average current through the device.  Therefore each extremum of
$U(N)$ corresponds to one of the branches of the $I$-$V$ curve.  Outside
the bistable region the current $I(V)$ is uniquely defined, and $U(N)$ has
a single minimum.  In the bistable region $U(N)$ has two minima and a
maximum, which correspond to the locally stable current branches and the
unstable branch, respectively [Fig.~\ref{fig:IVcurve}].

From the definitions of $A$ and $B$, it is clear that their ratio is
independent of the area of the sample $S$.  Using the
expression~(\ref{potential}) and the fact that $N\propto S$, one can see
that $U$ is linearly proportional to $S$.  Thus $U(N)$ is an extensive
quantity, and its dependence on $N$ and $S$ has the general form $U(N, S)
= S u(n)$, where $n=N/S$ is the electron density.  Since the area of the
sample is large, we have $U\gg 1$.  Therefore, the distribution function
$P_0$ is peaked sharply near the global minimum of $U(N)$.

\begin{figure}
  \resizebox{.36\textwidth}{!}{\includegraphics{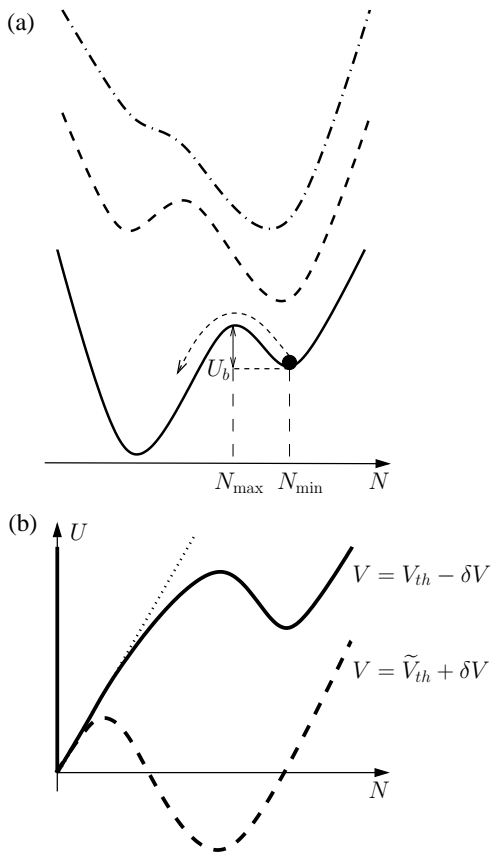}}
\caption{\label{fig:potential} (a) Generic behavior of $U(N)$ at different
  values of bias.  Outside the bistable region $U(N)$ has one minimum (top
  curve).  Inside the bistable region the function $U(N)$ has two minima
  and a maximum, which correspond to the locally stable current branches
  and the unstable branch, respectively (middle and bottom curves).  (b)
  The sketch of $U(N)$ for the model of Fig.~\ref{fig:IVcurve}.  Solid
  line corresponds to a bias slightly below $V_{th}$, whereas dashed line
  depicts $U(N)$ for the bias slightly above ${\widetilde V}_{th}$.}
\end{figure}

The experiments~\cite{Grahn96, Grahn98, Teitsworth} studying the switching
between the branches of the $I$-$V$ curve are set up as follows.  One
starts at $V < {\widetilde V}_{th}$, Fig.~\ref{fig:IVcurve}, where only
one value of current is possible.  In this case $U(N)$ has only one
minimum, as shown schematically by the dash-dotted line in
Fig.~\ref{fig:potential}a.  If we increase the bias up to some value
slightly above ${\widetilde V}_{th}$, the function $U(N)$ will acquire a
new minimum to the left of the old one, see the dashed line in
Fig.~\ref{fig:potential}a.  This corresponds to the appearance of the
lower current branch of the $I$-$V$ curve.  The new minimum is a local
one, and the main peak of the distribution function is still centered at
the old minimum.  Thus, the system remains on the upper branch of the
$I$-$V$ curve.  Further increasing $V$, we transform $U(N)$ to the shape
shown schematically by the solid line in Fig.~\ref{fig:potential}a.  Here
the right minimum of $U(N)$ is a local one, and if we leave the system in
this state for a sufficiently long time, it will eventually switch to the
left minimum.  To switch from the local minimum to the global one, the
system has to overcome the barrier of height $U_b$,
Fig.~\ref{fig:potential}a.  From the form of the distribution
function~(\ref{potential}) it is clear that this process takes a long time
$\tau\propto \exp (U_{b})$.  To perform the measurement of the switching
time from the upper to the lower current branch, one increases the bias to
the chosen value over a time short compared to $\tau$, and then waits
until the system switches to the lower branch.

At the threshold the right minimum of $U(N)$ disappears, and $U_b =0$.
Since $U(N)$ is an extensive quantity, $U_b$ grows rapidly when we move
from the threshold into the bistable region, and the switching time
becomes very long.  Therefore, in order for the switching to occur within
a reasonable timeframe, the system should be close to the threshold.

As the voltage approaches its threshold value, the maximum at
$N_{\text{max}}$ and the local minimum of $U(N)$ at $N_{\text{min}}$
(Fig.~\ref{fig:potential}) move closer to each other, and at the threshold
they coincide.  At this point one can define a threshold electron density
$n_{th}\equiv N_{\text{max}}/S = N_{\text{min}}/S$.  In the vicinity of
$n=n_{th}$ and $V=V_{th}$ the function $u(n)$ can be approximated by a
cubic polynomial,
\begin{equation}
\label{cubicPotential}
u(n)\approx -\alpha (n-n_{th}) + \frac{\gamma}{3}(n-n_{th})^3
+ u_{th},\quad \alpha =a(V_{th}-V).
\end{equation}
Here the constant $u_{th}$ is the value of $u$ at $n=n_{th}$ and
$V=V_{th}$.

To derive Eq.~(\ref{cubicPotential}) microscopically, one has to
consider $A(N)$ and $B(N)$ on the upper branch of the $I$-$V$ curve in
the vicinity of the threshold $V_{th}$.  An analytical calculation of
$I(V)$ is possible~\cite{Blanter99} if the dimensionless parameter
\begin{equation}
\label{smalllambda}
\lambda = \frac{me^2}{2\pi \hbar^{2}c}
\end{equation}
is small, $\lambda\ll 1$.  Here $c=C/S$ is the capacitance per unit
area.  In Appendix~\ref{appendix:AB} we extend this approach to find
$A(N)$, $B(N)$, and the coefficients of
expansion~(\ref{cubicPotential}) at $\lambda \ll 1$.

The expansion~(\ref{cubicPotential}) can be justified for any $\lambda$ in
the spirit of the Landau theory of second-order phase
transitions.~\cite{LandauStat1} The potential $u$ is expected to be an
analytic function of $n$ and $V$.  Thus, $u$ can be expanded in Taylor
series near the threshold, with $n- n_{th}$ playing the role of the order
parameter.  Since the local minimum and the maximum of $u$ coincide at the
threshold, both the first and second derivatives of $u(n)$ vanish at
$V=V_{th}$ and $n=n_{th}$.  Therefore, the expansion starts with the
third-order term.  The sign of $\gamma$ is not important; we choose
$\gamma >0$, which corresponds to the behavior of $U$ near the right
minimum as shown in Fig.~\ref{fig:potential}a.  At $V\neq V_{th}$, the
linear $\alpha (n -n_{th})$ and quadratic $\frac12 \beta (n -n_{th})^2$
terms are also present.  Since $\alpha =\beta =0$ at $V=V_{th}$, we expect
$\alpha \propto (V_{th}-V)$, $\beta \propto (V_{th} -V)$.  We keep only
the linear term in the expansion, because the second-order term is
quadratic in small parameter $n-n_{th}$, and therefore is small compared
to the linear one.  In order for $u$ to have a local minimum at $V<
V_{th}$, the coefficient $a$ should be positive.

Near the threshold the function $B(N)$ can be approximated by a
constant,
\begin{equation}
\label{Bconst}
B(N_{th}) = 2J_R = 2\Gamma_R N_{th}/\hbar.
\end{equation}
In the case of constant $B$ the Fokker-Planck
equation~(\ref{tunnelingFPE}) has been studied in detail.  In
particular, the exact expression for the mean switching time can be
obtained including the prefactor (Ref.~\onlinecite{vanKampen},
Sec.~XIII.2).  In our notations it reads
\begin{equation}
\label{Tau}
\tau = \frac{4\pi} {B \sqrt{U''(N_{\text{min}})
|U''(N_{\text{max}})|}} \exp (U_{b}).
\end{equation}

For the potential~(\ref{cubicPotential}) one can easily find the barrier
height, $U_b = 4S\alpha^{3/2}/3\gamma^{1/2}$.  The prefactor of
Eq.~(\ref{Tau}) can be also straightforwardly evaluated, and one obtains
the following expression for the mean switching time,
\begin{equation}
\label{Tau0}
\tau = \frac{2\pi }{b\sqrt{\alpha \gamma}}\exp\left[
\frac43\frac{S [a(V_{th}-V)]^{3/2}}{\gamma^{1/2}}\right],
\end{equation}
where $b=B/S$ is independent of the area of the sample.  This result
obviously agrees with Eq.~(\ref{paper1}) for small samples ($L\ll r_0$).

Expansion~(\ref{cubicPotential}) is quite generic, and similar theoretical
results were found in many different areas of physics.~\cite{Kurkijarvi,
  Victora, Dykman1, Dykman2, Dykman:exp, Dykman3, Dykman4} In particular,
Eq.~(\ref{tunnelingFPE}) is also used to describe the motion of a Brownian
particle in external potential, where $N$ plays the role of the coordinate
of the particle.  Therefore, the logarithm of the mean escape time of the
Brownian particle from a local minimum of potential is also expected to
obey the $3/2$-power law.  Recently this behavior of the escape time was
confirmed experimentally for the optically trapped Brownian
particle.~\cite{Dykman:exp}

The lower branch of the $I$-$V$ curve corresponds to the situation where
the level in the well $E_0 +e\phi$ is below the bottom $eV$ of the
conduction band in the left lead.  In this case $J_L \equiv 0$ and $B=
-A=J_R$.  Consequently, as $N\to 0$ we have $U(N) = 2N$, see
Eq.~(\ref{potential}).  Since $N$ cannot be negative, $U$ reaches its
minimum at the boundary $N=0$ of the range of allowed values of $N$, where
the derivative $U'(N)\neq 0$, Fig.~\ref{fig:potential}b.  The
non-analyticity of $U(N)$ near the left minimum does not affect the
calculation of the time of switching from upper to the lower branch.
Indeed, at a bias slightly below $V_{th}$, the function $U(N)$ is analytic
near its maximum and the local minimum (solid line in
Fig.~\ref{fig:potential}b), and the description of the switching from the
upper to the lower branch in terms of Eqs.~(\ref{cubicPotential})
and~(\ref{Tau}) is correct.  However, the situation is different for the
switching from the lower to the upper branch of the $I$-$V$ curve.  To
study this process, we decrease the bias to the value slightly above
${\widetilde V}_{th}$.  The function $U(N)$ for this case is depicted
schematically by the dashed line in Fig.~\ref{fig:potential}b.  Here
$U(N)$ is non-analytic at its local minimum, and therefore we cannot use
expressions~(\ref{cubicPotential}) and~(\ref{Tau}) for the switching time.
The non-analytic behavior of $U(N)$ is a consequence of crudeness of our
model, in which the current on the lower branch is exactly zero.  On the
other hand, the experimentally measured $I$-$V$ curves show non-zero
current on the lower branch.  Thus, in a more detailed model which
accounts for this non-zero current, the minimum corresponding to the lower
branch of the $I$-$V$ curve will be reached at non-zero $N$.  The
discussion based on Eqs.~(\ref{cubicPotential}) and~(\ref{Tau}) will then
be valid.

\section{\label{sec:3} Fokker-Planck equation for transport in DBRTS of
  large area}

In section~\ref{sec:2} we studied the decay of a metastable state in DBRTS
under the assumption that the electron density in the quantum well is
uniform.  Then the switching time $\tau$ given by Eq.~(\ref{Tau0}) grows
exponentially with the area of the sample.  Since electrons can tunnel at
any point of the quantum well, the tunneling process creates a non-uniform
electron density.  On the other hand, the diffusion of particles in the
well leads to spreading of the charge across the sample.  In small samples
the spreading is fast, and the density becomes uniform.  In samples of
large area the electron density may change significantly before the charge
spreads over the entire well.  In this case the switching between the two
branches of the $I$-$V$ curve is initiated in a small part of the sample,
and the switching time is not exponential in the area $S$.

In this section we generalize the Fokker-Planck
equation~(\ref{tunnelingFPE}) to the case of non-uniform density ${n({\bf
r})}$, where $\textbf{r}=(x,y)$ is a position in the well.  In subsequent
sections this equation will be used to study the decay of a metastable
state in DBRTS of large area.

\subsection{\label{diffusion} Equation for distribution function of
  electron density in an isolated quantum well}

We begin by considering the simplest case of a quantum well not coupled to
the leads.  At finite temperature the electron density in the well
fluctuates and can be described by a distribution function
$P\{n(\mathbf{r}),t\}$.  Here we derive the Fokker-Planck equation for the
distribution function of electron density due to the in-plane diffusion of
electrons in the well.  In section~\ref{sec:FPEcombined} we add the
tunneling through the barriers and obtain the Fokker-Planck equation for
DBRTS of large area.

We consider density fluctuations at length scales much greater than the
inelastic mean free path.  These density fluctuations are slow in
comparison with the energy relaxation time in the well.  Therefore the
system is in a local equilibrium, and the distribution of electrons at any
point in the well is given by a Fermi function.  Note that the chemical
potential in this Fermi function is determined by the electron density,
and therefore varies from point to point following the dependence
$n(\mathbf{r})$.

Let us choose a time interval $\Delta t$ much smaller than the relaxation
time for $P\{n(\mathbf{r}),t\}$ and large in comparison with the collision
time, so that the motion of electrons can be treated as diffusive.  Then
one can write the following equation for the evolution of the distribution
function,
\begin{eqnarray}
\label{discreteP}
&& P\{n(\mathbf{r}),t+\Delta t\} - P\{n(\mathbf{r}),t\}
= \int\!\!\int  d\mathbf{r}_1  d\mathbf{r}_2
\nonumber \\
&&\times \Big[
P\{ n(\mathbf{r}) + \delta n_{12}(\mathbf{r}),t\}
W_{\Delta t}(\mathbf{r}_1 ,\mathbf{r}_2 ;n(\mathbf{r})
+ \delta n_{12}(\mathbf{r}))
\nonumber \\
&& - P\{n(\mathbf{r}),t\}
W_{\Delta t}(\mathbf{r}_2 ,{\bf r}_1 ;n(\mathbf{r})) \Big].
\end{eqnarray}
Here $\delta n_{12}({\bf r}) = \delta({\bf r}-{\bf r}_1) - \delta({\bf
r}-{\bf r}_2)$ is the correction to the density $n(\mathbf{r})$ due to the
displacement of one electron from point $\mathbf{r}_2$ to $\mathbf{r}_1$;
the probability density $W_{\Delta t}({\bf r}_1 ,{\bf r}_2
;n(\mathbf{r}))$ describes diffusion of an electron from a point ${\bf
r}_1$ in the quantum well to point ${\bf r}_2$ during the time interval
$\Delta t$.  Since the diffusion rate may depend on the electron density,
$W_{\Delta t}$ is a functional of $n(\mathbf{r})$.

Expanding the first term in the right-hand side of Eq.~(\ref{discreteP})
up to the second order in $\delta n_{12}({\bf r})$, one obtains the
following equation:
\begin{eqnarray}
\label{FPE1}
\Delta P\{n,t\}
& = &\frac{1}{2} \int\!\!\int d{\bf r}_1 d{\bf r}_2
\left\{\left(\frac{\delta}{\delta n({\bf r}_1)}
- \frac{\delta}{\delta n({\bf r}_2)}\right)
\right.
\nonumber \\
&&\times [ W_{\Delta t}({\bf r}_1 ,{\bf r}_2 ;n)
- W_{\Delta t}({\bf r}_2 ,{\bf r}_1 ;n) ]
\nonumber \\
&&+ \frac{1}{2} \left( \frac{\delta}{\delta n({\bf r}_1)}
- \frac{\delta}{\delta n({\bf r}_2)} \right)^{2}
[ W_{\Delta t}({\bf r}_1 ,{\bf r}_2 ;n)
\nonumber \\
&&+ W_{\Delta t}({\bf r}_2 ,{\bf r}_1 ;n) ]\bigg\} P\{n,t\}.
\end{eqnarray}
The probability densities $W_{\Delta t}$ to diffuse from $\mathbf{r}_1$ to
$\mathbf{r}_2$ and back are not independent,
\begin{equation}
\label{detailed}
W_{\Delta t}({\bf r}_1 ,{\bf r}_2 ;n) e^{-\mu_{1}/T}
= W_{\Delta t}({\bf r}_2 ,{\bf r}_1 ;n) e^{-\mu_{2}/T}.
\end{equation}
Here $\mu_{1}$ and $\mu_{2}$ are the electrochemical potentials at points
$\mathbf{r}_1$ and ${\bf r}_2$, respectively.  For the case of elastic
scattering by impurities considered in Ref.~\onlinecite{paper1}
expression~(\ref{detailed}) directly follows from Eq.~(8) of
Ref.~\onlinecite{paper1}.  Generalization of Eq.~(\ref{detailed}) to
arbitrary scattering mechanism is discussed in
Appendix~\ref{appendix:balance}.

In order to find $\mu$ we need to account for the interactions between
electrons.  We limit ourselves to the charging energy approximation; the
electron exchange and correlation effects are neglected.  Then at low
temperatures $T\ll E_F$, the values of the electrochemical potential are
found by adding the electrostatic potential $e^{2}n/c$ to the Fermi
energy,
\begin{equation}
\label{mu1,2}
\mu_{1,2} = \frac{e^2}{\tilde c}n({\bf r}_{1,2}).
\end{equation}
Here the effective capacitance per unit area $\tilde c$ is defined by
$e^{2}/{\tilde c} = e^{2}/{c} + 1/{\nu}$, and $\nu$ is the density of
states in the well.

In short time $\Delta t$ an electron can only diffuse over a short
distance, so that $|\mu_1-\mu_2 |\ll T$.  Therefore using
Eq.~(\ref{detailed}), one can expand the expression in the curly brackets
in the right-hand side of Eq.~(\ref{FPE1}) to the leading order in
$(\mu_1-\mu_2 )/T$, and with the help of Eq.~(\ref{mu1,2}) obtain
\begin{eqnarray}
\label{FPE2}
\Delta P\{n,t\}
& = &\frac{1}{2} \int\!\!\int\!\! d{\bf r}_1 d{\bf r}_2
\Bigg\{ \frac{e^2}{{\tilde c}T}
\left(\frac{\delta}{\delta n({\bf r}_1)}
- \frac{\delta}{\delta n({\bf r}_2)}\right)
\nonumber \\
&&\!\left. \times [n({\bf r}_{1}) - n({\bf r}_{2})]
+ \left(\frac{\delta}{\delta n({\bf r}_1)}
- \frac{\delta}{\delta n({\bf r}_2)}\right)^{2} \right\}
\nonumber \\
&&\times
W_{\Delta t}({\bf r}_1 ,{\bf r}_2 ;n) P\{n,t\}.
\end{eqnarray}

To proceed further we need an expression for the transition probability
density $W_{\Delta t}$.  This quantity is affected by all the relevant
processes of electron scattering, such as elastic scattering of electrons
by impurities, electron-phonon and electron-electron scattering.  Instead
of accounting for all these processes explicitly, we express $W_{\Delta
t}$ in terms of in-plane conductivity $\sigma$, which can in principle be
measured experimentally.  Assuming that electron motion is diffusive, we
conclude that the average square of the distance traveled by an electron
during a short time interval is proportional to $\Delta t$, i.e.,
\begin{equation}
\label{Gdeltat}
\int\!\! W_{\Delta t}({\bf r}_1 ,{\bf r}_2 ;n)
|{\bf r}_{1} - {\bf r}_2 |^2 \, d{\bf r}_2 = G \Delta t.
\end{equation}
Here the constant $G$ is proportional to the conductivity, $G = 4T\sigma
/e^2$, see Appendix~\ref{appendix:derivationG}.

At small $\Delta t$ the transition probability density $W_{\Delta t}$ can
be expanded as
\begin{equation}
\label{approxW}
W_{\Delta t}({\bf r}_1 ,{\bf r}_2 ;n)
= \delta({\bf r}_1 -{\bf r}_2 )
+ \frac{T\sigma \Delta t}{e^2}\nabla^2
\delta ({\bf r}_1 -{\bf r}_2 ) +\dots~.
\end{equation}
The physical meaning of the first term in this expansion is that electron
remains at its initial position $\mathbf{r}_1$ at $\Delta t =0$.  Thus the
second term is needed to account for the electron diffusion.  The
coefficient in the second term is found by applying the
expansion~(\ref{approxW}) to Eq.~(\ref{Gdeltat}).

Equation~(\ref{FPE2}) can be simplified significantly using
expansion~(\ref{approxW}), and eventually takes the form
\begin{equation}
\label{inplaneFPE}
\frac{\partial}{\partial t} P\{n,t\} = -\frac{\sigma}{e^2}
\int\!\! d{\bf r}\, \frac{\delta}{\delta n}
\left[ \frac{e^2}{\tilde c} \nabla^2 n
+ T\nabla^2 \frac{\delta}{\delta n} \right] P\{ n,t\}.
\end{equation}
This is the Fokker-Planck equation for the evolution of the distribution
function of electron density.  The first term in Eq.~(\ref{inplaneFPE})
describes the spreading of the charge in the well, whereas the second term
accounts for the thermal noise.

It is instructive to substitute into Eq.~(\ref{inplaneFPE}) the
equilibrium distribution function $P_0 \{n\}$.  The latter has the Gibbs
form $e^{-E/T}$, namely,
\begin{equation*}
P_0 \{n\} = \exp\left[ -\frac{1}{T}\int
\frac{e^{2}n^2({\bf r})}{2\tilde c}\, d{\bf r}\right].
\end{equation*}
Here the energy per unit area $\varepsilon = {e^{2}n^2} /{2{\tilde c}}$ is
chosen in a way that reproduces the electrochemical potential $\mu =
\partial \varepsilon /\partial n$ in the form (\ref{mu1,2}).  It is easy
to check that $P_0 \{n\}$ satisfies the Fokker-Planck
equation~(\ref{inplaneFPE}).

\subsection {\label{sec:FPEcombined} Combined Fokker-Planck equation for
  tunneling and diffusion}

In this section we obtain the combined Fokker-Planck equation which
incorporates both the tunneling through the barriers and diffusion inside
the well.  We begin by generalizing the tunneling Fokker-Planck
equation~(\ref{tunnelingFPE}) to the case of non-uniform electron density.
This is accomplished by dividing the plane of the well into small pieces,
so that the density is uniform within each piece.  In the absence of
in-plane diffusion, the distribution function of electron density in the
entire plane is given by the product of distribution functions of its
pieces, $P = \prod_j P_j \{N_j \}$.  Applying Eq.~(\ref{tunnelingFPE}) to
each piece we obtain the following Fokker-Planck equation for the
distribution function of the entire quantum well,
\begin{equation*}
\frac{\partial}{\partial t} P
= \sum_j \frac{\partial }{\partial N_j} \left[- A(N_j )
+ \frac{1}{2}\frac{\partial}{\partial N_j}B(N_j ) \right] P.
\end{equation*}

The functions $A(N_j)$ and $B(N_j)$ are extensive quantities, and it is
convenient to rewrite them as $A(N_j )=\Delta S a(n)$ and $B(N_j )=\Delta
S b(n)$, where $\Delta S$ is the area of each piece.  Replacing the sum
with the integral over the area of the sample and $\partial /\partial N_j$
with the functional derivative $\delta/\delta n(\mathbf{r}_j)$, we find
the continuous form of this equation:
\begin{equation}
\label{continuousFPE}
\frac{\partial}{\partial t} P\{n,t\}
= \!\int\!\! d \mathbf{r} \frac{\delta}{\delta n}
\left[ -a(n(\mathbf{r})) + \frac{1}{2} \frac{\delta}{\delta n}
b(n(\mathbf{r})) \right] P\{n,t\}.
\end{equation}

Let us now take into account the in-plane diffusion of electrons, which
was discussed in Sec.~\ref{diffusion}.  Because the tunneling and
diffusion are independent processes, we can add the right-hand sides of
Eqs.~(\ref{continuousFPE}) and~(\ref{inplaneFPE}) and obtain the combined
Fokker-Planck equation for DBRTS of large area:
\begin{eqnarray}
\label{FPEwithAllTerms}
\frac{\partial}{\partial t} P\{n,t\}
& = &\int\!\! d \mathbf{r}\,\frac{\delta}{\delta n} \left[ -a(n)
+ \frac{1}{2} \frac{\delta}{\delta n} b(n)
\right.
\nonumber \\
&&\left. - \frac{\sigma}{\tilde c} \nabla^{2}n
- T\frac{\sigma}{e^2} \nabla^{2} \frac{\delta}{\delta n} \right]
P\{ n,t\}.
\end{eqnarray}
This equation generalizes Eq.~(\ref{inplaneFPE}) to the case of a quantum
well coupled to the leads.

In the vicinity of the threshold $V_{th}$ the function $b(n)$ can be
approximated by a constant $b= b(n_{th})$.  In addition, one can
substitute $2a/b = 2A/B = -u'(n)$, c.f. Eq.~(\ref{potential}).  At bias
near $V_{th}$ the function $u(n)$ is given by the approximate expression
(\ref{cubicPotential}), and Eq.~(\ref{FPEwithAllTerms}) can be rewritten
as
\begin{eqnarray}
\label{FPEgeneral}
\frac{\partial}{\partial t} P\{n,t\} & = &\frac{b}{2}
\int\!\! d \mathbf{r}\, \frac{\delta}{\delta n}
\bigg[ -\alpha + \gamma(n-n_{th})^{2}
\nonumber \\
&&\left. - 2\eta\nabla^{2}n + \frac{\delta}{\delta n} \right]
P\{ n,t\},
\end{eqnarray}
where we defined $\eta = \sigma /{\tilde c}b$.  In Eq.~(\ref{FPEgeneral})
we omitted the term proportional to the temperature, since it is
negligible at low $T$.  (The exact criterion is discussed in
Appendix~\ref{appendix:tempTerm}.)  Thus from now on we study only the
effect of the shot noise due to the tunneling of electrons at high bias
$eV\gg T$, whereas the thermal noise is neglected.

The stationary solution of Eq.~(\ref{FPEgeneral}) is found by setting the
left-hand side to zero,
\begin{eqnarray}
\label{P0general}
P_{0}\{n\} & = & e^{-F\{n\}},
\nonumber \\
F\{n\}
& = &\int\!\! d{\bf r} \left[ - \alpha(n-n_{th})
+ \frac{\gamma}{3} (n-n_{th})^{3} + \eta (\nabla n)^{2} \right].
\nonumber \\
\end{eqnarray}
The functional $F\{n\}$ has two contributions: the first two terms account
for the tunneling, and the remaining term is due to the in-plane
diffusion.

The functional $F\{n\}$ is similar to the free energy in Ginzburg-Landau
theory of phase transitions, with $n-n_{th}$ playing the role of the order
parameter.  For the case of uniform electron density in the well, $\nabla
n = 0$, the functional $F\{n\}$ coincides with Eq.~(\ref{potential}).  If
the density is non-uniform, the gradient term $(\nabla n)^{2}$ appears in
the expansion in addition to the terms from Eq.~(\ref{potential}).  This
gradient term suppresses large variations of the electron density.

\subsection{\label{sec:dimensionless} Dimensionless Fokker-Planck equation}

For the following discussion it is convenient to parametrize the electron
density $n(\mathbf{r})$ in terms of a dimensionless function
$z(\bm{\rho})$ that vanishes at the minimum of $u(n)$,
\begin{subequations}
\begin{eqnarray}
\label{nr}
n({\bf r})& = &n_{min} -2\sqrt{\frac{\alpha}{\gamma}}\, z({\bf r}/r_{0}),\\
\label{r0}
r_0& = &\frac{\sqrt{\eta}}{(\alpha\gamma)^{1/4}} =
\sqrt{\frac{\sigma}{{\tilde c} b \sqrt{\alpha\gamma}}}.
\end{eqnarray}
\end{subequations}
Here the density at the minimum $n_{min} = n_{th} + \sqrt{\alpha /\gamma}$
can be easily found from Eq.~(\ref{cubicPotential}).  The Fokker-Planck
equation~(\ref{FPEgeneral}) in terms of $z(\bm{\rho})$ takes the form:
\begin{eqnarray}
\label{dimensionlessFPE}
\frac{\partial P\{z,t\}}{\partial t}& = &b\sqrt{\gamma\alpha}
\int\!\! d\bm{\rho}\, \frac{\delta}{\delta z}
\left[ -\nabla_{\rho}^{2}z + z - z^{2}
+ \frac{1}{U_0} \frac{\delta}{\delta z} \right]
\nonumber \\
&&\times P\{z,t\},
\end{eqnarray}
where
\begin{equation}
\label{U2}
U_0 = \frac{8\eta\alpha}{\gamma}.
\end{equation}
The stationary solution $P_{0}$ of Eq.~(\ref{dimensionlessFPE}) is given
by
\begin{equation}
\label{P0}
P_{0}\{z\} = e^{-F},\qquad F = U_0 \int\!\! d\bm{\rho}
\left[\frac{(\nabla_{\rho}z)^{2}}{2} + \frac{z^{2}}{2}
- \frac{z^{3}}{3}\right].
\end{equation}

One can see that the characteristic value of the functional $F$ is given
by $U_0$, whereas the characteristic size $r_0$ plays the role of a
typical length scale of stochastic fluctuations of electron density
$n(\mathbf{r})$.

\section {\label{sec:decay} Decay of the metastable state in extended samples}

In Sec.~\ref{sec:2} we obtained the expression for the mean switching time
in DBRTS under the assumption of uniform electron density in the well.
This assumption is valid only if the dimensions of the sample are small
compared to the length scale $r_0$ of the density fluctuations,
Eq.~(\ref{r0}).  If the sample is large, the fluctuations of electron
density must be taken into account.

In Sec.~\ref{sec:3} we obtained the Fokker-Plank
equation~(\ref{FPEgeneral}) which describes the time evolution of the
distribution function of electron density.  Unlike
Eq.~(\ref{tunnelingFPE}) for the case of uniform density, this equation
has an infinite number of variables, since the density is different at
every point.

The most general form of the multidimensional Fokker-Planck equation is
\begin{eqnarray}
\label{L}
\!\!\!\!\!\frac{\partial P({\bf x},t)}{\partial t}& = & {\cal L}\,
P({\bf x},t),
\nonumber \\
{\cal L}& = &-\sum\limits_{i}\frac{\partial }{\partial x_{j}}
K_{i} (\mathbf{x})
+ \sum\limits_{i,j}\frac{\partial^2}{\partial x_{i} \partial x_{j}}
D_{ij}({\bf x}).
\end{eqnarray}
Assuming that the system has a metastable state, one can consider its
domain of attraction $\Omega$.  The domain boundary $\partial\Omega$ is a
separatrix of the drift field $\mathbf{K}$.  The mean time of the first
passage out of the domain $\Omega$ has been found in
Refs.~\onlinecite{Matkowsky, Hanggi}.  For the process described by
Eq.~(\ref{L}) the mean switching time is obtained as doubled mean
first-passage time~\cite{Hanggi} and takes the form,
\begin{equation}
\label{Hanggi}
\tau = - \frac{2\int_{\Omega}d^{d}x\,
P_0 ({\bf x})}{\sum\limits_{i} \int_{\partial \Omega} dS_{i}\,
\sum\limits_{j} D_{ij}({\bf x})P_0 ({\bf x})
\frac{\partial f({\bf x})}{\partial x_{j}}}.
\end{equation}
Here $P_0$ is the stationary solution of Eq.~(\ref{L}).  The form function
$f({\bf x})$ is a stationary solution of the adjoint equation,
\begin{equation}
\label{f}
{\cal L}^{\dagger}f({\bf x},t) = \sum\limits_{j}\left(K_{j}({\bf x})
+ \sum\limits_{i} D_{ij}({\bf x}) \frac{\partial}{\partial x_{i}}\right)
\frac{\partial f({\bf x})}{\partial x_{j}} = 0.
\end{equation}
In addition, $f({\bf x})$ is defined to vanish at the boundary
$\partial\Omega$ and reach $f({\bf x}) \simeq 1$ well inside $\Omega$.

In subsequent sections we use the expression~(\ref{Hanggi}) to find the
mean time of current switching in double-barrier structures.

\subsection{\label{small} Mean switching time in small samples}

In samples with linear dimensions small compared with $r_0$ the density
fluctuations are weak.  In this section we study their effect on the mean
switching time.  We will show that even these weak fluctuations can result
in significant change of $\tau$.

\subsubsection{\label{sec:small1} Evaluation of the mean switching time}

In order to bring the Fokker-Planck equation~(\ref{dimensionlessFPE}) to
the form~(\ref{L}) we present $z(\bm{\rho})$ as an expansion
\begin{equation}
\label{x}
z(\bm{\rho}) = \sum\limits_{i =0}^{\infty} x_i \phi_{i}(\bm{\rho}),
\end{equation}
where $\phi_{i}(\bm{\rho})$ are the normalized eigenfunctions of the
Laplace operator, $-\nabla_{\rho}^{2} \phi_{i}(\bm{\rho}) = \epsilon_i
\phi_{i}(\bm{\rho})$.  In particular $\phi_{0}(\bm{\rho}) = r_0 /\sqrt{S}$
and $\epsilon_0 =0$.  Since there is no current flowing through the
boundaries of the sample, the eigenfunctions must satisfy the boundary
conditions $\hat{\bm n}\cdot \bm{\nabla} \phi_{i} (\bm{\rho})
\big|_{\rm{boundary}}=0$, where $\hat{\bm n}$ is a unit vector normal to
the boundary.  The $x_0$-coordinate corresponds to the average electron
density in the well, whereas the other coordinates describe small
fluctuations of the density.  The eigenvalues $\epsilon_i$ are numbered in
order of increasing magnitude; $\epsilon_1 \sim r_0^2 /S \gg 1$.

To obtain the $\mathbf{x}$-representation of the Fokker-Planck equation we
substitute the expansion~(\ref{x}) into Eq.~(\ref{dimensionlessFPE}) and
find
\begin{eqnarray}
\label{XrepFPE}
\mathcal{L} &=&
b\sqrt{\gamma\alpha} \left\{ \sum_{i=0}^{\infty}
\frac{\partial}{\partial x_i}
\left[ (\epsilon_i + 1 -2\phi_0 x_0 ) x_i
+ \frac{1}{U_0 } \frac{\partial}{\partial x_i} \right] \right.
\nonumber \\
&&+ \frac{\partial}{\partial x_0} \phi_0 x_0^2
+ \sum_{i,j,k=1}^{\infty} \xi_{ijk}
\frac{\partial}{\partial x_i} x_j x_k \Bigg\},
\end{eqnarray}
where
\begin{equation*}
\xi_{ijk} = \int\!\! d\bm{\rho}\, \phi_{i}(\bm{\rho}) \phi_{j}(\bm{\rho})
\phi_{k}(\bm{\rho}).
\end{equation*}

The stationary solution $P_0 = e^{-F}$ in terms of $x_i$ can be found by
substituting expression~(\ref{x}) into Eq.~(\ref{P0}). Then the functional
$F$ takes the form,
\begin{eqnarray}
\label{Px}
F\{\mathbf{x}\}& = & U_0 \left[ \frac{1}{2}
\sum\limits_{i=0}^{\infty} (\epsilon_i + 1 -2\phi_0 x_0 ) x_{i}^{2}
+ \frac{2 \phi_0 x_0^3}{3} \right.
\nonumber \\
&&- \frac{1}{3} \sum_{i,j,k=1}^{\infty} \xi_{ijk} x_i x_j x_k
\Bigg].
\end{eqnarray}
One can easily verify that $\exp(-F\{\mathbf{x}\})$ solves the
Fokker-Planck equation with $\mathcal{L}$ given by Eq.~(\ref{XrepFPE}).

The stationary probability density $P_0$ is sharply peaked at the minimum
of the functional $F$, i.e., at $z (\bm{\rho}) = 0$ ($\mathbf{x} =0$).
Therefore, keeping terms up to the second order in $x_i$ in
expansion~(\ref{Px}), we can evaluate the integral in the numerator of
Eq.~(\ref{Hanggi}) in Gaussian approximation:
\begin{equation}
\label{numerator}
\int_{-\infty}^{\infty} \prod_{i=0}^{\infty} dx_i \, P\{\mathbf{x}\}
= \prod_{i=0}^{\infty} \sqrt{\frac{2\pi}{U_0 ( \epsilon_i + 1)}}.
\end{equation}

In a multidimensional case in order to switch from the metastable state
the system has to pass from the local minimum of $F$ to its global
minimum.  The switching process is dominated by the paths which go through
the vicinity of the lowest saddle point separating the domains of
attraction of metastable and stable states.  The boundary of the domain
$\Omega$ lies exactly at the saddle point and is orthogonal to the
direction of the steepest descent.

The integral in the denominator of Eq.~(\ref{Hanggi}) is dominated by the
saddle point of $F$.  The latter is found from the condition $\delta
F/\delta z =0$.  This equation has an obvious solution $z_s
(\bm{\rho})=1$.  In $\mathbf{x}$-representation it corresponds to $x_0 =
1/\phi_0$ and $x_i =0$ for $i\ge 1$.  Expanding expression~(\ref{Px}) near
this point up to the second order in $x_i$ we approximate $F$ near the
saddle point by
\begin{equation}
\label{Ps}
F\{\mathbf{x}\}
\simeq U_0 \left[ \frac{1}{6\phi_0^2}
- \frac{1}{2} \left( x_0 - \frac{1}{\phi_0} \right)^2
+ \frac{1}{2} \sum_{i=1}^{\infty} (\epsilon_i - 1) x_i^2 \right].
\end{equation}
In small samples $\epsilon_1 >1$, and therefore $F$ has only one unstable
direction $x_0$, whereas all other directions are stable.  One can see
from Eq.~(\ref{Ps}) that in this approximation the boundary $\partial
\Omega$ is the plane $x_0 = 1/\phi_0$.

Since the boundary $\partial \Omega$ is orthogonal to the $x_0$-direction,
the sum over $i$ in the denominator of Eq.~(\ref{Hanggi}) reduces to a
single term with $i=0$.  Comparing Eqs.~(\ref{L}) and~(\ref{XrepFPE}) one
finds that $D_{ij} = (b\sqrt{\gamma\alpha} /U_0 ) \delta_{ij}$.  Noting
that $D_{ij}$ is diagonal, the sum over $j$ also reduces to the only term
with $j=0$.

To find $\partial f /\partial x_0$ one needs to solve Eq.~(\ref{f}).
Noting that $\epsilon_0 = 0$ and using Eq.~(\ref{XrepFPE}), we can write
the adjoint equation~(\ref{f}) near the saddle point as
\begin{equation}
\left[U_0  \left( x_0 - \frac{1}{\phi_0} \right)
+ \frac{\partial}{\partial x_{0}}\right]
\frac{\partial f}{\partial x_{0}} = 0.
\end{equation}
Solving this equation, we obtain
\begin{equation}
\label{derivativef}
\frac{\partial f}{\partial x_{0}}
= - \sqrt{\frac{2U_0 }{\pi}}\,
e^{-\frac{U_0}{2} \left( x_0 - \frac{1}{\phi_0} \right)^2}.
\end{equation}
Here the prefactor was found using the fact that $f=1$ inside the domain
$\Omega$ (i.e., at $x_0 \to -\infty$) and $f=0$ at the domain boundary
$x_0 = 1/\phi_0$.

Using Eqs.~(\ref{Ps}) and (\ref{derivativef}) we can evaluate the integral
in the denominator of Eq.~(\ref{Hanggi}) in Gaussian approximation.  Then
dividing the numerator~(\ref{numerator}) by this integral, we find the
following expression for the mean switching time,
\begin{equation}
\label{tauexp}
{\tilde \tau} = \tau \Upsilon_0 .
\end{equation}
Here $\tau$ is the switching time~(\ref{Tau0}) obtained without
the inclusion of density fluctuations.  The latter give rise to
the renormalization factor
\begin{equation}
\label{ups0}
\Upsilon_0 =
\prod\limits_{i=1}^{\infty}
\sqrt{\frac{\epsilon_i -1}{\epsilon_i +1}}.
\end{equation}

To estimate the product $\Upsilon_0$ we assume a rectangular geometry of
the sample with length $L$ and width $w$.  Then the eigenvalues
$\epsilon_i$ are given by
\begin{equation}
\label{lambda}
\epsilon_i = \epsilon_{nm}
= \pi^2 r_0^2 \left(\frac{m^2}{L^2} + \frac{n^2}{w^2}\right),
\end{equation}
where $n,m$ are non-negative integers.

In small samples $\epsilon_{nm}\gg 1$ and the expression for
$\ln\Upsilon_0$ can be expanded as
\begin{equation}
\label{sum}
\ln\Upsilon_0 \simeq - \sump\limits_{n,m=0}\frac{1}{\epsilon_{nm}}
= - \frac{1}{\pi^2 r_0^2} \sump\limits_{n,m=0}
\frac{1}{\frac{m^2}{L^2} + \frac{n^2}{w^2}},
\end{equation}
where the prime in the sum means that the term with $n=m=0$ is excluded.

The infinite sum in Eq.~(\ref{sum}) is logarithmically divergent.
However, since the diffusion picture is only valid at distances greater
than the mean free path $l$, the wavevectors of the density fluctuations
cannot\footnote{At wavevectors larger than $1/l$ the motion of electrons
  becomes ballistic, and therefore the conductivity $\sigma\to\infty$.
  Then it follows from the expression~(\ref{r0}) that $r_0 \to\infty$, and
  thus the sum~(\ref{sum}) is cut off at wavevectors of the order of
  $1/l$.} exceed $l^{-1}$.  Therefore, we need to cut the sum off at $m\le
L /l$ and $n\le w /l$.

At $L\sim w\sim\sqrt{S}$ the sum~(\ref{sum}) can be approximated by a
two-dimensional integral and yields,
\begin{equation}
\label{logUpsilon0}
\ln \Upsilon_0
\simeq - \frac{S}{2\pi r_0^2} \ln\frac{\sqrt{S}}{l}.
\end{equation}
Note, that although in small samples the area $S$ is small
compared to $r_0^2$, the effect of density fluctuations may become
significant at $l\ll \sqrt{S}$.

In the case of strip geometry, $w\ll L$, we separate the sum into
two parts, with $n=0$ and $n>0$.  The first part gives the sum of
$1/m^2$ which can be explicitly evaluated and results in a small
contribution $L^2 /6r_0^2 \ll 1$ to $\ln\Upsilon_0$.  In the
second part we approximate the sum over $m$ by the integral with
an infinite upper limit.  Then neglecting terms $\sim (L/r_0)^2$,
we obtain the sum of $1/n$.  Cutting off this sum as discussed
above, we find
\begin{equation}
\ln \Upsilon_0
\simeq - \frac{Lw}{2\pi r_0^2} \ln\frac{w}{l}.
\end{equation}

For simplicity, from now on we will consider samples with $w\sim
L\sim\sqrt{S}$.

\subsubsection{\label{sec:renormalization} Renormalization of threshold voltage}

Using Eqs.~(\ref{tauexp}),~(\ref{logUpsilon0}) and~(\ref{Tau0}) we
find the following expression for the mean switching time in small
samples,
\begin{eqnarray}
\label{tauwithlog}
{\tilde \tau} &=& \frac{2\pi}{b\sqrt{\alpha
\gamma}} \exp\left[ \frac43\frac{S [a(V_{th} -
V)]^{3/2}}{\gamma^{1/2}} \right.
\nonumber \\
&& - \frac{S\sqrt{\gamma a}(V_{th} - V)^{1/2}}{2\pi \eta}
\ln\frac{\sqrt{S}}{l} \bigg].
\end{eqnarray}
The second term in the exponential of Eq.~(\ref{tauwithlog}) represents
the correction~(\ref{logUpsilon0}) due to the density fluctuations.

Let us consider the regime when the magnitude of this term is larger
than unity, but still small compared to the first term in the
exponential of Eq.~(\ref{tauwithlog}).  Then this correction can be
interpreted as a shift of the threshold voltage in
formula~(\ref{Tau0}).  Indeed, substituting $V_{th}\to V_{th} + \delta
V_{th}$, with the shift
\begin{equation}
\label{deltaV}
\delta V_{th} = - \frac{1}{4\pi} \frac{\gamma}{a \eta}
\ln\frac{\sqrt{S}}{l},
\end{equation}
into Eq.~(\ref{Tau0}) and expanding it up to the first order in $\delta
V_{th}$ we reproduce the result~(\ref{tauwithlog}).  In experiments the
threshold voltage $V_{th}$ is not known \textit{a priori}.  If one treats
it as a fitting parameter, Eqs.~(\ref{tauwithlog}) and~(\ref{Tau0}) are
equivalent up to the first order in $\delta V_{th}$.

The last term in the exponential of Eq.~(\ref{tauwithlog}) formally
diverges at $l\to 0$.  Similar divergences have been studied in quantum
field theory in the problem of the decay of the false
vacuum.~\cite{Voloshin:lecture, Voloshin, Coleman1, Coleman2, Coleman3,
  ZinnJustin, Selivanov} According to Eq.~(\ref{cubicPotential}), the
shift~(\ref{deltaV}) of the threshold voltage is equivalent to adding a
linear in the order parameter term $-a\delta V_{th} (n - n_{th})$ to the
integrand of the functional~(\ref{P0general}).  This corresponds to the
standard in quantum field theory method of renormalization of action,
Refs.~\onlinecite{Voloshin:lecture, Voloshin, Coleman1, Coleman2,
  Coleman3, ZinnJustin, Selivanov}.  Such renormalization procedure
removes all the divergences.

The origin of the renormalization of the threshold voltage can be
understood as follows.  The ``action'' $F$ describes the so called
$\phi^3$ field theory in two dimensions, where $\phi \equiv (n - n_{th})$
is a scalar field.  An alternative approach to the renormalization of this
scalar field theory is to integrate out the fast modes $\phi_F$
corresponding to large wavevectors, while keeping only slow modes $\phi_S$
with small wavevectors in the action $F$.  One can find that the averaging
of $\phi_F^2$ gives the sum of inverse eigenvalues of Laplace operator
identical to~(\ref{sum}), so that the term $\frac{\gamma}{3}\phi^3$ after
the integration over the fluctuations of the fast modes gives rise to
$\gamma \langle \phi_F^2 \rangle \phi_S = - a\delta V_{th} \phi_S$.
Physically this renormalization corresponds to the averaging of the
switching rate over fluctuations of the electron density $n$ in the well
with characteristic scales between the mean free path and the sample size.

Due to the renormalization of the threshold voltage the parameter $\alpha$
is modified as $\alpha \to\alpha + a\delta V_{th}$.  Therefore, the
quantities which depend on $\alpha$, such that $\tau$ and $r_0$, are also
renormalized.  More precise expression for $\tau$ is given by
Eq.~(\ref{Tau0}) upon substitution of the renormalized $\alpha$ into it.
On the other hand, the small corrections to the prefactor of $\tau$ due to
the renormalization are more challenging to observe experimentally, and
for comparison with experiment they can be ignored.

\section {\label{sec:large} Mean switching time in large samples}

So far we studied samples of small area $S \ll r_0^2$.  We found that the
switching occurs when the electron density at the saddle point is uniform,
because the diffusion processes are fast and they smooth out all density
variations.  In large samples, $S \gg r_0^2$, the diffusion is slower, and
the system can reach the critical density in a small part of the well.
After the switching occurs in that part, the switching process spreads
rapidly throughout the entire well.  In this section we study the
switching time due to these nucleation processes.

To find the expression for the mean switching time $\tau$ in large samples
we need to obtain the minimum and the saddle points of the functional $F$
in Eq.~(\ref{P0}).  They can be found using the condition $\delta F/\delta
z = 0$, i.e.,
\begin{equation}
\label{kink}
- \nabla^{2} z +z -z^2=0.
\end{equation}
The boundary conditions for Eq.~(\ref{kink}) should account for the fact
that there is no current flowing through the boundaries of the sample.
Since the current is proportional to the density gradient $\bm{\nabla} n$,
according to Eq.~(\ref{nr}) these boundary conditions take the form
\begin{equation}
\label{bc}
\hat{\bm n}\cdot \bm{\nabla} z \big|_{\rm{boundary}}=0,
\end{equation}
where $\hat{\bm n}$ is a unit vector normal to the boundary. The trivial
solution $z(\bm{\rho})=0$ gives the minimum of the functional $F\{z\}$,
while the saddle points can be found as non-trivial solutions
$z_{s}(\bm{\rho})$ of Eq.~(\ref{kink}).

\subsection{\label{large1} Nucleation processes in very large samples}

\begin{figure}
  \resizebox{.41\textwidth}{!}{\includegraphics{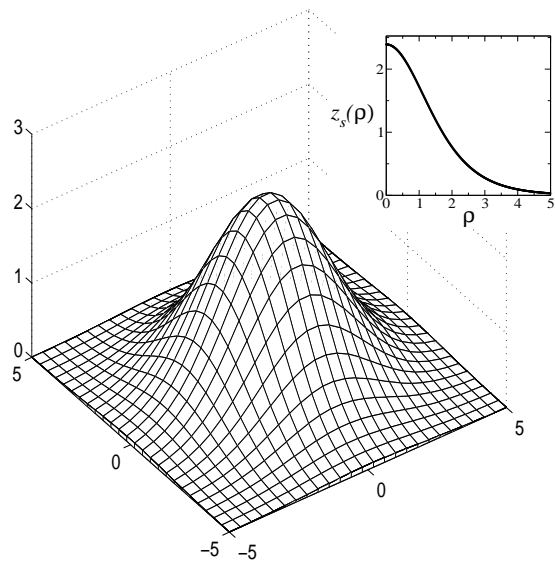}}
\caption{\label{fig3}The sketch of the density profile at the saddle point
  $z_s (\bm{\rho})$ corresponding to the solution of Eq.~(\ref{kink}) with
  the boundary conditions~(\ref{bc}).  The precise radial dependence $z_s
  (\rho)$ obtained by solving Eq.~(\ref{zradial}) numerically is shown in
  the inset. }
\end{figure}

Let us consider the switching in an infinite sample, $S\to \infty$.  Due
to the symmetry of the problem, the solutions of Eq.~(\ref{kink}) should
be azimuthally symmetric. Placing the origin of coordinate system at the
center of switching region and writing Eq.~(\ref{kink}) in polar
coordinates, we find
\begin{equation}
\label{zradial}
z''_s (\rho) + \frac{1}{\rho}z'_s (\rho) - z_s (\rho)
+ z_s^2 (\rho) = 0.
\end{equation}
This equation should be solved with the boundary condition $z_s (\rho)=0$
at $\rho\rightarrow \infty$, since otherwise $F\{z_s \} \propto S$, and
the switching time $\tau\propto e^{F\{z_s \}}$ will be infinite at $S\to
\infty$.  One can show that this condition is consistent with
Eq.~(\ref{bc}), that is $z'_s (\infty)=0$.  Indeed, Eq.~(\ref{kink}) can
be interpreted as a Schr\"odinger equation for a particle in potential
$-z_s$, i.e., $- (\nabla^{2} +z_s)z_s =-z_s$.  Therefore, $z_s (\rho)$ has
the meaning of an eigenfunction of a bound state; its asymptotic behavior
at large distances is $z_s\to e^{-\rho}/\sqrt{\rho}$, so that $z'_s
(\infty)=0$.  The non-trivial solution of Eq.~(\ref{zradial}) with the
boundary condition described earlier can be obtained numerically.  The
result is shown in the inset of Fig.~\ref{fig3}.

The main exponential dependence of mean switching time $\tau_i$ in an
infinite sample is given by $e^{F\{z_s \}}$.  Substituting the numerical
result for $z_s(\bm{\rho})$ into Eq.~(\ref{P0}), one finds \cite{paper1}
\begin{equation}
\label{exptaularge}
\tau_i = \tau_i^* \exp\left[ 8\zeta\, \frac{\eta
a(V_{th}-V)}{\gamma} \right],
\end{equation}
where the numerical constant
\begin{equation}
\label{eq:zeta}
\zeta = \int\!\! d\bm{\rho} \left[\frac{(\nabla
z_s)^2}{2} + \frac{z_s^2}{2} -\frac{z_s^3}{3} \right] \approx 7.751.
\end{equation}

Equation~(\ref{exptaularge}) is the counterpart of the result~(\ref{Tau0})
derived for small samples, $S\ll r_0^2$.  Because of the dependence of
$r_0$ on $V$, see Eqs.~(\ref{r0}) and~(\ref{cubicPotential}), both types
of behavior can be observed in a single device by tuning the bias.  At the
crossover, $r_0^2 \sim S$, the two results coincide.

The problem of stochastic current switching is similar to the problem of
finding the probability of spontaneous decay of a metastable vacuum near a
Peierls transition point in ($1+1$) dimensional scalar field theory.  The
latter problem was solved in Ref.~\onlinecite{Selivanov}, and the
exponential factor in the result for the decay time is analogous to
Eq.~(\ref{exptaularge}).  On the other hand, the prefactor of the decay
time is essentially different from $\tau_i^*$, since we study the shot noise
described by classical Fokker-Planck equation, while the false vacuum
decay problem is inherently quantum mechanical.

\subsubsection{\label{sec:prefactor}Evaluation of the prefactor}

In a finite sample the switching can occur anywhere in the well, hence the
prefactor of the switching rate $\tau_i^{-1}$ must be proportional to the
area $S$.  Thus, while $\tau_i$ has a large exponential,
Eq.~(\ref{exptaularge}), its prefactor $\tau_i^*$ is proportional to $1/S$
and can be small in large samples. Therefore, to fully understand the
switching one needs to find $\tau_i^*$.

The time evolution of distribution function $P\{z,t\}$ in large samples is
given by the Fokker-Planck equation~(\ref{dimensionlessFPE}).  To evaluate
the prefactor of the mean switching time we again use the
expression~(\ref{Hanggi}).  The procedure is similar to the one for small
samples described in Sec.~\ref{small}.  However, the integration in
Eq.~(\ref{dimensionlessFPE}) is now over a large sample, and therefore the
density at the saddle point becomes non-uniform, Fig.~\ref{fig3}.  This
significantly complicates the evaluation of the prefactor $\tau_i^*$.

We evaluate both integrals in Eq.~(\ref{Hanggi}) in gaussian
approximation.  As in Sec.~\ref{small} the integral in the numerator of
Eq.~(\ref{Hanggi}) is dominated by the minimum of $F\{z\}$ and is given by
the expression~(\ref{numerator}).  The denominator of~(\ref{Hanggi}) is
dominated by the saddle point.  Presenting $z(\bm{\rho})$ near the saddle
point as $z(\bm{\rho}) = z_{s}(\rho) +{\tilde v}(\bm{\rho})$, we obtain
the expansion of $F\{z\}$ in the form,
\begin{equation}
\label{Fwithv}
F\{z_s + {\tilde v}\} = F\{z_s \}
+\frac{U_0 }{2}\!\int\!\! d\bm{\rho}\, {\tilde v}(\bm{\rho})
[ -\nabla_{\rho}^{2} - 2z_{s} +1] {\tilde v}(\bm{\rho}).
\end{equation}

It is convenient to evaluate the integral in Eq.~(\ref{Fwithv}) by
expanding
\begin{equation}
\label{tildev}
{\tilde v}(\bm{\rho}) = \sum\limits_{n,m}{\tilde x}_{nm}{\tilde
  \phi}_{nm}(\bm{\rho}),
\end{equation}
where ${\tilde \phi}_{nm}(\bm{\rho})$ are the normalized solutions of the
eigenvalue problem
\begin{equation}
\label{2Dequation}
[-\nabla_{\rho}^{2} -2z_{s}(\bm{\rho}) +1] {\tilde \phi}_{nm}(\bm{\rho})
= {\tilde \lambda}_{nm} {\tilde \phi}_{nm}(\bm{\rho}).
\end{equation}
The boundary conditions for this equation are given by Eq.~(\ref{bc}).

Equation~(\ref{2Dequation}) can be interpreted as a Schr\"odinger equation
for a particle in the attractive potential $-2z_s$ with energy ${\tilde
\lambda}_{nm} - 1$.  Since the potential is azimuthally symmetric, we can
separate the variables as ${\tilde \phi}_{nm}(\bm{\rho}) = Q_{nm}(q\rho)
\Psi_{m}(\varphi)$.  The solutions for the azimuthal part
$\Psi_{m}(\varphi)$ are given by $e^{\pm im\varphi}$.  Below it will be
convenient to use their real combinations, $\cos m\varphi$ and $\sin
m\varphi$, and introduce the following notations: $\Psi_{0}(\varphi) =
1/\sqrt{2\pi}$, $\Psi_{m}(\varphi) = (\cos m\varphi)/\sqrt{\pi} $ for
$m=1,2,\dots$, and $\Psi_{m}(\varphi) = (\sin m\varphi)/\sqrt{\pi}$ for
$m=-1,-2,\dots$.

Substituting the expansion~(\ref{tildev}) into Eq.~(\ref{Fwithv}) and
using the orthonormality condition for the eigenfunctions, we find
\begin{equation}
\label{largePs}
F\{\mathbf{{\tilde x}}\} = F\{z_s \} + \frac{U_0}{2}
\sum\limits_{n=0}^{\infty}\sum\limits_{m=-\infty}^{\infty} {\tilde
\lambda}_{nm}{\tilde x}_{nm}^{2}.
\end{equation}

The discussion leading to Eq.~(\ref{largePs}) did not rely on the
assumption of large sample size.  In the case of small samples
Eq.~(\ref{largePs}) reproduces the expansion~(\ref{Ps}), if one identifies
${\tilde \lambda} = \epsilon -1$.  This relation is easily understood by
noticing that in small samples the density at the saddle point is $z_s
(\bm{\rho}) =1$.  Comparing the definition of $\epsilon$ given in the
paragraph after Eq.~(\ref{x}) with Eq.~(\ref{2Dequation}), where we
substitute $z_s =1$, we reproduce ${\tilde \lambda} = \epsilon -1$.

The form of Eq.~(\ref{largePs}) suggests that in the case of large samples
it is more convenient to evaluate the integral in the denominator of
Eq.~(\ref{Hanggi}) using variables ${\tilde x}_{nm}$ rather than $x_{nm}$.
Since the eigenfunctions $\phi_{nm}$ and ${\tilde \phi}_{nm}$ are
normalized, the expansion coefficients ${\tilde x}_{nm}$ are related to
coefficients $x_{nm}$ of expansion~(\ref{x}) via an orthogonal
transformation.  The Jacobian of this transformation equals unity, and
therefore the integration over $\prod dx_{nm}$ in the denominator of
Eq.~(\ref{Hanggi}) can be replaced by the integration over $\prod d{\tilde
  x}_{nm}$.

In order to evaluate the integral in the denominator of Eq.~(\ref{Hanggi})
in the $\mathbf{{\tilde x}}$-representation, we need to find the
eigenvalues ${\tilde \lambda}_{nm}$ of Eq.~(\ref{2Dequation}).  All
${\tilde \lambda}_{nm}$ are positive with the exception of one negative
eigenvalue, ${\tilde \lambda}_{00} <0$, and two zero eigenvalues, ${\tilde
  \lambda}_{0,1} = {\tilde \lambda}_{0,-1}=0$.  Numerical solution of
equation~(\ref{2Dequation}) yields ${\tilde\lambda}_{00} \approx -1.648$.
This negative eigenvalue is associated with unstable deviation from $z_s$
corresponding to the motion over the saddle point.  In Eq.~(\ref{Hanggi})
the boundary $\partial\Omega$ of the domain of attraction of the
metastable state is orthogonal to ${\tilde x}_{00}$-direction, so that the
integration in the denominator is performed only over the positive and
zero modes. Since each positive ${\tilde\lambda}_{nm}$ corresponds to a
Gaussian integral, the integration over them is straightforward. The
integration over the zero modes is more challenging; to perform it we
first need to understand their physical meaning.

The existence of two zero eigenvalues is due to the translational
invariance of the functional $F\{z\}$ with respect to any shift of the
center of the switching region in the plane of the quantum well.  The two
zero eigenvalues correspond to two orthogonal to each other directions in
the plane along which such a shift can be performed.  Indeed, a small
shift $\Delta\bm{\rho}$ of the center of switching region results in the
following small change in the saddle point density,
\begin{equation}
\label{shiftRho}
z_s (\bm{\rho} +\Delta\bm{\rho} ) - z_s (\bm{\rho}) =
\frac{\partial z_s}{\partial\rho_x}\Delta\rho_x
+ \frac{\partial z_s}{\partial\rho_y}\Delta\rho_y.
\end{equation}
One can check by differentiating Eq.~(\ref{kink}) with respect to
$\rho_{x,y}$ that the derivatives $\partial z_s /\partial \rho_{x,y}$ are
solutions of Eq.~(\ref{2Dequation}) with ${\tilde\lambda} =0$.
Furthermore, $\partial z_s /\partial \rho_{x} = z'_s (\rho) \cos\varphi$
and $\partial z_s /\partial \rho_{y} = z'_s (\rho) \sin\varphi$, so the
azimuthal quantum numbers corresponding to zero modes are $m=\pm 1$ in our
notations.  Thus we conclude that $\partial z_s /\partial \rho_{x,y} = c_0
{\tilde \phi}_{0,\pm 1} (\bm{\rho})$, where $c_0$ is a
constant.\footnote{For the zero modes the quantum number $n=0$, because
  the radial part of the wavefunction $z'_s (\rho)$ has no zeros at finite
  $\rho$, and thus corresponds to the ground state at $m=\pm 1$.}

Substituting these expressions for $\partial z_s /\partial \rho_{x,y}$
into Eq.~(\ref{shiftRho}) and comparing it with the
expansion~(\ref{tildev}), we find that the coefficients corresponding to
zero modes are ${\tilde x}_{0,\pm 1} = c_0 \Delta\rho_{x,y}$.  Thus the
integral over the zero modes ${\tilde x}_{0,1}$ and ${\tilde x}_{0,-1}$
amounts to the integration over the possible positions of the center of
the switching region in the sample,
\begin{equation}
\label{zeromodes}
\int\!\! d{\tilde x}_{0,1}\int\!\! d{\tilde x}_{0,-1} =
c_0^2 \int\! d(\Delta\rho_x ) \int\! d(\Delta\rho_y ) =
\zeta \frac{S}{r_0^2}.
\end{equation}
Here the constant $c_0$ was found using azimuthal symmetry of $z_s$ and
the fact that the eigenfunctions ${\tilde \phi_{nm}}(\bm{\rho})$ are
normalized,
\begin{equation}
\label{czeromode}
c_0^2 = \int\!\! \left( \frac{\partial z_s}{\partial\rho_x} \right)^2
\, d\bm{\rho}
= \frac12 \int\! [\nabla z_{s}(\bm{\rho})]^2 \, d\bm{\rho} = \zeta.
\end{equation}
The relation between the last integral and the constant $\zeta$ defined by
Eq.~(\ref{eq:zeta}) is proven in Appendix~\ref{appendix:zeta}.

To find the denominator of Eq.~(\ref{Hanggi}) in the $\mathbf{{\tilde
    x}}$-representation we also need $D_{ij}$ and $\partial f/\partial
{\tilde x}_{00}$.  They can be obtained from the $\mathbf{{\tilde
    x}}$-representation of the Fokker-Planck equation for large samples.
Substituting $z(\bm{\rho}) = z_{s}(\rho) +{\tilde v}(\bm{\rho})$ with
${\tilde v}$ in the form~(\ref{tildev}) into Eq.~(\ref{dimensionlessFPE})
and using the orthonormality condition for the eigenfunctions ${\tilde
  \phi}_{nm} (\bm{\rho})$, we obtain the Fokker-Planck equation
$\dot{P}={\cal L}P$ with
\begin{equation}
\label{Xrepresentation}
{\cal L} = b\sqrt{\gamma\alpha} \sum_{n,m}
\left[ \frac{\partial}{\partial {\tilde x}_{nm}}
{\tilde \lambda}_{nm}{\tilde x}_{nm}
+ \frac{1}{U_0}\frac{\partial^2}{\partial {\tilde x}_{nm}^{2}} \right].
\end{equation}
Here we neglected the terms of higher orders in ${\tilde x}_{nm}$.  One
can easily check that the solution of the stationary Fokker-Planck
equation ${\cal L} P_0 =0$ is $P_0 = e^{-F}$ with $F$ given by the Gaussian
approximation~(\ref{largePs}).

Comparing Eqs.~(\ref{L}) and~(\ref{Xrepresentation}) we conclude that
$D_{ij} = (b\sqrt{\gamma\alpha} /U_0 ) \delta_{ij}$.  To find $\partial f
/\partial {\tilde x}_{00}$ we need to solve Eq.~(\ref{f}) with ${\cal L}$
given by~(\ref{Xrepresentation}), that is
\begin{equation}
\label{eq:X}
\left[ -{\tilde \lambda}_{00}{\tilde x}_{00}
+ \frac{1}{U_0}\frac{\partial}{\partial {\tilde x}_{00}} \right]
\frac{\partial f}{\partial {\tilde x}_{00}} = 0.
\end{equation}
Solving it with the conditions $f=1$ inside the domain $\Omega$ (i.e., at
$x_0 \to -\infty$) and $f=0$ at the domain boundary $x_{00} = 0$, we find
that at the saddle point $\partial f /\partial {\tilde x}_{00} = - (2
|{\tilde\lambda}_{00}|U_0 /\pi)^{1/2}$.

Substituting Eq.~(\ref{numerator}) for the numerator of
Eq.~(\ref{Hanggi}), and Eqs.~(\ref{largePs}) and (\ref{zeromodes}) along
with the expressions for $D_{ij}$ and $\partial f/\partial {\tilde
  x}_{00}$ into the denominator of Eq.~(\ref{Hanggi}) we reproduce the
result~(\ref{exptaularge}) with the prefactor given by
\begin{equation}
\label{2}
\tau_i^* \simeq \frac{\pi^{2}}{4\sqrt{|{\tilde
\lambda}_{00}|} \zeta bS\alpha^{2}}
\Upsilon''',\quad
\Upsilon''' = {\prod_{n,m}}'''
\sqrt{\frac{{\tilde \lambda}_{nm}}{\lambda_{nm}} }.
\end{equation}
Here the product $\Upsilon'''$ excludes the factors corresponding to the
three non-positive eigenvalues ${\tilde\lambda}_{nm}$.  The coefficients
$\lambda_{nm}$ denote the parameters $1 + \epsilon_{i}$ used in
Sec.~\ref{small}.  They coincide with the eigenvalues of the Schr\"odinger
equation~(\ref{2Dequation}) in the absence of the attractive potential
$-2z_s (\rho)$.

To evaluate the infinite product $\Upsilon'''$ we need to find the
continuous spectrum of Eq.~(\ref{2Dequation}).  The radial part of
${\tilde\phi}_{nm}$ oscillates as a function of $\rho$ with the wavevector
$q_n$.  The phase of these oscillations at $\rho\to\infty$ is shifted by
$\delta_m (q_n )$ due to the scattering in the attractive potential
$-2z_s$.  The eigenvalues of the continuous spectrum are expressed in
terms of these scattering phase shifts as follows
\begin{equation}
\label{tillambdanm}
{\tilde \lambda}_{nm} = 1 + q_{n}^{2}
\simeq 1 +\left(\frac{\pi n}{R}\right)^{2}
\left[1 - \frac{\delta_{m}(\pi n/R)}{\pi n}\right]^{2}.
\end{equation}
This result is derived for a round sample of dimensionless radius $R\gg
1$; the derivation and the expression for the phase shifts $\delta_{m}$
are given in Appendix~\ref{appendix3}.  The expression for the eigenvalues
$\lambda_{nm}$ is given by Eq.~(\ref{tillambdanm}) with $\delta_{m} = 0$.

It is convenient to calculate the logarithm of $\Upsilon'''$, thereby
transforming the product over $n$ and $m$ to a sum.  Taking the large
sample limit, $R\to\infty$, we replace the sum over $n$ by an integral
over $q=\pi n/R$. Then expanding the integrand in small parameter
$\delta_m /n$, we find
\begin{equation}
\label{upsint}
\ln\Upsilon''' \simeq  - \frac{1}{\pi} \int_{0}^{\infty}
\left[ \sum\limits_{m=-\infty}^{\infty}
\delta_{m}(q)\right] \frac{q\, dq}{1+q^{2}}.
\end{equation}

To investigate the convergence of the integral we need to evaluate the sum
of the phase shifts at large $q$.  This is accomplished with the help of
the following ``Friedel sum rule''
\begin{equation}
\label{7}
\left. \sum\limits_{m=-\infty}^{\infty}
\delta_{m}(q) \right|_{q\to\infty}
= 2\zeta
\end{equation}
proven in Appendix~\ref{appendix3}.  The asymptotic
behavior~(\ref{7}) of the phase shifts implies that the integral
in Eq.~(\ref{upsint}) diverges logarithmically at $q\to\infty$.
This ultraviolet divergence signals that $\Upsilon'''$ is
determined by a large wavevector cutoff or, equivalently, by some
short distance scale.  An analogous divergence appeared in the
prefactor of the mean switching time in small samples,
Sec.~\ref{small}. There we have shown that this short distance
cutoff is of the order of the mean free path $l$.  Following the
same recipe, we cut off the integral in Eq.~(\ref{upsint}) at $q
\sim r_{0}/l$, and with logarithmic accuracy find
\begin{equation}
\label{largesum}
\ln\Upsilon''' \simeq  -\frac{2\zeta}{\pi}
\ln\left(\frac{r_0}{l}\right).
\end{equation}
Substituting this result into Eq.~(\ref{2}), we obtain the prefactor
\begin{equation}
\label{prefactor}
\tau_i^* \sim \frac{1} {bS\alpha^{2}}
\left( \frac{l}{r_0}\right)^{2\zeta /\pi}.
\end{equation}
This expression completely describes the parametric dependence of the
prefactor of the mean switching time in large samples.  On the other hand,
because of the ultraviolet divergence of $\Upsilon'''$, the numerical
coefficient in $\tau_i^*$ cannot be determined without detailed treatment
of charge transport at short length scales.~\footnote{Apart from the
  continuous spectrum, equation~(\ref{2Dequation}) has one discrete
  positive eigenvalue ${\tilde\lambda}_{10}\approx 0.771$. It gives rise
  to a factor $\sqrt{{\tilde\lambda}_{10}}$ in $\Upsilon'''$ that was not
  accounted for in Eq.~(\ref{largesum}).  We neglect this factor along
  with the unknown numerical coefficient in Eq.~(\ref{prefactor}).}

\subsubsection{\label{renormalization2} Renormalization of threshold
voltage in large samples}

Expression~(\ref{prefactor}) for the prefactor $\tau_i^*$ implies that in
large samples the switching rate $\tau_i^{-1}$ diverges at $l\to 0$.  In
Sec.~\ref{sec:renormalization} we encountered the same problem while
considering small samples. There it was shown that the dependence of
$\tau$ on the mean free path can be absorbed into the definition of the
threshold voltage $V_{th}$.  Following the same renormalization technique,
one can shift the threshold voltage $V_{th}$ by the amount
\begin{equation}
\label{newdeltaV}
\delta V_{th} = -\frac{1}{4\pi} \frac{\gamma}{a\eta}
\ln\frac{r_0}{l},
\end{equation}
chosen in such a way that the resulting correction to the exponential in
Eq.~(\ref{exptaularge}) cancels $\Upsilon'''$ in the prefactor $\tau_i^*$,
see Eqs.~(\ref{2}),~(\ref{largesum}).  The renormalized result for the
mean switching time then takes the form
\begin{equation}
\label{tau_i}
\tau_i \sim \frac{1}{bS\alpha_R^{2}}
\exp\left[ \frac{8\zeta\eta\alpha_R}{\gamma} \right],\quad
\alpha_R =  \alpha + a\delta V_{th}.
\end{equation}
This expression is equivalent to Eqs.~(\ref{exptaularge}),
(\ref{prefactor}) up to the correction in the prefactor $\tau_i$ due to
the substitution $\alpha\to\alpha_R$.

The characteristic length scale $r_0$ is sensitive to the position of the
threshold voltage, so its value has to be renormalized.  Since the size of
the critical nucleus and $\delta V_{th}$ are coupled to each other,
Eq.~(\ref{newdeltaV}), they should be evaluated self-consistently:
\begin{subequations}
\label{systemr0Vth}
\begin{eqnarray}
\label{newr0withR}
r_R &=&  \frac{\sqrt{\eta}}{[ (\alpha + a\delta V_{th})\gamma ]^{1/4}},\\
\label{realnewdeltaV}
\delta V_{th} &=& -\frac{1}{4\pi} \frac{\gamma}{a\eta}
\ln\frac{r_R}{l}.
\end{eqnarray}
\end{subequations}
To find $\delta V_{th}$ one can solve the system of
equations~(\ref{systemr0Vth}) iteratively starting with $r_R =r_0$.  The
result~(\ref{newdeltaV}) then should be understood as the first iteration
of Eq.~(\ref{realnewdeltaV}).

Upon the substitution of the shift~(\ref{realnewdeltaV}) into
Eq.~(\ref{tau_i}), the logarithm of the switching time $\tau_i$ acquires
an additional logarithmic dependence on voltage due to the bias-dependent
renormalization of $V_{th}$.  This dependence is physically meaningful and
can, in principle, be tested experimentally.  However, these corrections
to the voltage dependence~(\ref{exptaularge}) of $\tau_i$ are small, and
to the leading order $\ln \tau_i$ is still linear in voltage.

\subsection{Nucleation near sample boundaries}

In Sec.~\ref{large1} we studied the nucleation processes in very large
samples assuming that the switching initiates far from the boundaries
(i.e., at distances significantly greater than $r_0$).  In this section we
show that the switching can be more effective when it is initiated near
the boundaries of the sample and evaluate the mean switching time for such
processes.

\subsubsection{\label{large2} Nucleation at a smooth edge}

To study the nucleation near an edge which is smooth on the scale $r_0$,
we model the sample by a half-plane and set up the coordinate system so
that $\rho_x$ is the coordinate along the boundary and $\rho_y$ is
positive inside the half-plane.  Then the boundary condition~(\ref{bc})
takes the form $z'_{\rho_y}(\rho_x , +0) =0$.  If we place the center of
the saddle-point solution $z_s$ shown in Fig.~\ref{fig3} on the edge of
the sample, the resulting function
\begin{equation}
\label{edgesolution} z_e (\rho_x,\rho_y) = z_s ( [(\rho_x
-\rho_x^{(0)})^2 +\rho_y^2]^{1/2})
\end{equation}
automatically satisfies not only equation~(\ref{kink}) but also the
boundary condition. Therefore, the expression~(\ref{edgesolution}) gives
the saddle point density for the half-plane.

One can argue that there are no other saddle point solutions for edge
switching.  Indeed, suppose that we have a solution ${\tilde z}(\rho_x
,\rho_y)$ of Eq.~(\ref{kink}) for a half-plane.  Then we can define
function $z(\rho_x ,\rho_y)$ in the entire plane, so that $z={\tilde
  z}(\rho_x ,\rho_y)$ for $\rho_y >0$, and $z={\tilde z}(\rho_x ,-\rho_y)$
for $\rho_y <0$. By construction $z(\rho_x ,\rho_y)$ satisfies
Eq.~(\ref{kink}) at $\rho_y \neq 0$. However, this procedure does not
guarantee that the derivative $z'_{\rho_y}$ is continuous at $\rho_y =0$;
as a result $\partial^2 z/\partial \rho_y^2$ may have a delta-function
contribution.  More specifically, $z(\rho_x ,\rho_y)$ satisfies the
equation
\begin{equation}
\label{boundaryjump} -\nabla^2 z + z -z^2 = -2{\tilde
z}'_{\rho_y}(\rho_x , +0)\delta(\rho_y).
\end{equation}
If in addition ${\tilde z}(\rho_x ,\rho_y)$ satisfies the boundary
condition~(\ref{bc}), i.e., ${\tilde z}'_{\rho_y}(\rho_x , +0) =0$,
equation~(\ref{boundaryjump}) coincides with Eq.~(\ref{kink}) everywhere
in the plane. Then by construction $z(\rho_x, \rho_y) =z_s (\rho)$, and
therefore ${\tilde z}(\rho_x ,\rho_y)$ is given by a half of the
saddle-point solution $z_s$ shown in Fig.~\ref{fig3} with its center on
the boundary of the half-plane.  Thus, there are no saddle point solutions
for edge switching except~(\ref{edgesolution}).

The main exponential dependence of the mean switching time $\tau$ is given
by $e^{F\{z_s\}}$. In the definition~(\ref{P0}) of $F\{z\}$ the integral
is taken over the area of the sample.  In the case of switching far from
the boundaries it is over an entire plane, while for the edge switching
this integral is over a half-plane.  Therefore $F$ is reduced by a factor
of $2$ compared to the case of switching far from the boundaries.  Thus,
instead of Eq.~(\ref{exptaularge}), the expression for $\tau$ at the edge
takes the form
\begin{equation}
\label{exptauedge} 
\tau_e = \tau_e^* \exp\left[ 4\zeta\, 
\frac{\eta a(V_{th}-V)}{\gamma} \right].
\end{equation}

The evaluation of the prefactor $\tau_e^*$ is similar to the one for the
switching in the middle of a large sample, Sec.~\ref{sec:prefactor}.  In
that case we found two types of modes for the azimuthal part of the
eigenfunctions ${\tilde \phi}_{nm}(\bm{\rho})$ of Eq.~(\ref{2Dequation}),
namely, $\sin m\varphi$ and $\cos m\varphi$.  At the edge only the
eigenfunctions proportional to $\cos m\varphi$ are consistent with the
boundary condition $z'_{\rho_y}(\rho_x , +0) =0$ on the dimensionless
density $z$.  In the notations of Sec.~\ref{sec:prefactor} these modes
correspond to $m=0, 1, 2,\dots$.

The functional $F\{z\}$ is invariant with respect to the shifts of density
$z(\rho_x,\rho_y)$ along the edge of the sample.  Thus $F\{z\}$ has a
single zero mode ${\tilde x}_{01}$; it corresponds to the eigenfunction
with the azimuthal part $\cos \varphi$.  Integration over the zero mode,
in analogy with Eq.~(\ref{zeromodes}), is performed as
\begin{equation}
\label{zeromode1} \int d{\tilde x}_{01} =
\sqrt{\frac{\zeta}{2}}\frac{{\mathcal P}}{r_{0}},
\end{equation}
where ${\mathcal P}$ is the perimeter of the sample.

To evaluate the prefactor we again use formula~(\ref{Hanggi}).
Expression~(\ref{numerator}) for the numerator and the formulas for
$D_{ij}$ and $\partial f/\partial {\tilde x}_{00}$ in large samples are
still applicable, as they were obtained in a way independent of the exact
form of the saddle-point density.  Following the procedure of
Sec.~\ref{sec:prefactor}, we find the prefactor $\tau_e^*$ in the form
\begin{equation}
\label{22} 
\tau_e^* = \frac{\pi^{3/2} \Upsilon''}{\sqrt{2\zeta
|{\tilde \lambda}_{00}|} b \gamma^{1/4}\alpha^{5/4} {\mathcal
P}},\quad \Upsilon'' = \prodpp_{n,m\geq 0}
\sqrt{\frac{{\tilde \lambda}_{nm}}{\lambda_{nm}}},
\end{equation}
c.f. Eq.~(\ref{2}).  The definition of $\Upsilon''$ assumes that the
factors corresponding to the two lowest eigenvalues, ${\tilde
  \lambda}_{00}$ and ${\tilde \lambda}_{01}$, are excluded.

In the product $\Upsilon''$ the quantum number $m$ changes from $0$ to
$\infty$, while in $\Upsilon'''$ the same product is from $-\infty$ to
$\infty$.  Thus using the fact that ${\tilde\lambda_{nm}}$ and
$\lambda_{nm}$ are even functions of $m$, we obtain
\begin{equation}
\label{largesum2} \ln \Upsilon'' \simeq \frac{1}{2} \ln\Upsilon'''
\simeq -\frac{\zeta}{\pi} \ln \left( \frac{r_0}{l}\right),
\end{equation}
see Eq.~(\ref{largesum}).

Similarly to Eq.~(\ref{prefactor}) we find the prefactor of $\tau$ for the
edge switching
\begin{equation}
\label{tau2log2}
\tau_e^* \sim \frac{1}{b {\mathcal P} \gamma^{1/4}\alpha^{5/4} } 
\left( \frac{l}{r_0} \right)^{\zeta/\pi}.
\end{equation}
Note that due to the ultraviolet divergence of $\Upsilon''$ we can
evaluate $\tau_e^*$ only up to an undetermined constant.

Performing the same renormalization~(\ref{newdeltaV}) of the threshold
voltage as in Sec.~\ref{renormalization2}, one can eliminate the explicit
dependence $l^{\zeta /\pi}$ of the prefactor on the mean free path and
obtain the following expression for the mean switching time at the edge
\begin{equation}
\label{tau_e}
\tau_e \sim \frac{1}{b {\mathcal P} \gamma^{1/4} \alpha_R^{5/4} }
\exp\left[\frac{4\zeta\eta\alpha_R}{\gamma}\right].
\end{equation}
The exponent in Eq.~(\ref{tau_e}) is factor of 2 smaller than the exponent
of $\tau$ for the switching far from the boundaries, Eq.~(\ref{tau_i}).
Far from the threshold the exponential factor is dominant, and therefore
edge switching is more efficient.  To determine which switching mechanism
is more efficient near the threshold, one needs to take into account the
dependences of the prefactors in Eqs.~(\ref{tau_i}) and~(\ref{tau_e}) on
the dimensions of the device.

\subsubsection{\label{large3} Nucleation in a corner}

In Sec.~\ref{large2} we considered the processes of switching initiated
near a smooth edge of the sample.  In samples with pronounced corners,
such as the devices of square or triangular shape, there is also a
possibility of nucleation in a corner.  As we will show, such processes
may be more efficient than the nucleation in the interior and at the edges
of the sample.

We consider a corner of angle $\theta <\pi$.  Similarly to the discussion
in the beginning of Sec.~\ref{large2}, one can show that the saddle-point
solution $z_s (\bm{\rho})$ centered at the corner both solves the
equation~(\ref{kink}) and satisfies the boundary condition~(\ref{bc}).

The subsequent consideration is similar to the one for the switching at a
smooth edge of a large sample, Sec.~\ref{large2}.  At $\theta < \pi$ the
functional $F\{z\}$ does not possess translational symmetry with respect
to the shifts of $z(\bm{\rho})$, and therefore there are no zero modes.
Due to the boundary condition~(\ref{bc}) the allowed modes of the
azimuthal part of the eigenfunction ${\tilde \phi}_{nm}(\bm{\rho})$ of
Eq.~(\ref{2Dequation}) are $\cos(\pi m\varphi /\theta)$.  Then instead of
Eqs.~(\ref{exptauedge}) and~(\ref{22}), we obtain
\begin{equation}
\label{23} 
\tau_c = \frac{2\pi \Upsilon'}{
\sqrt{|{\tilde\lambda}_{00}|} b\sqrt{\gamma\alpha}}
\exp\left[ \frac{4\theta\zeta\eta\alpha}{\pi\gamma} \right],\quad
\Upsilon' = \prodp_{n,m\geq 0} \sqrt{ \frac{{\tilde
\lambda}_{nm}}{\lambda_{nm}}}.
\end{equation}
Here ${\tilde \lambda}_{nm}$ are the eigenvalues of Eq.~(\ref{2Dequation})
with the boundary conditions~(\ref{bc}), which take the form
$(\partial{\tilde \phi}_{nm}/\partial\varphi) |_{\varphi=0,\theta} =0$ for
the corner switching.  Unlike in Secs.~\ref{sec:prefactor}
and~\ref{large2}, here at $n\gg 1$ the eigenvalues ${\tilde \lambda}_{nm}$
are given by Eqs.~(\ref{tillambdanm}),~(\ref{phaseshifts}) with $m$
replaced by $\pi m /\theta$.  The product $\Upsilon'$ excludes the factor
corresponding to the negative eigenvalue ${\tilde \lambda}_{00}$.

Following closely the calculations of Secs.~\ref{sec:prefactor}
and~\ref{large2}, one can find the prefactor of $\tau_c$, and the
expression for the mean switching time takes the form
\begin{equation}
\label{tau2log3} 
\tau_c \sim \frac{1}{b\sqrt{\gamma\alpha}} \left(
\frac{l}{r_0} \right)^{\theta\zeta/\pi^{2}} \exp\left[
\frac{4\theta\zeta\eta\alpha}{\pi\gamma} \right].
\end{equation}

One might expect that at $\theta\to\pi$ this result should coincide with
Eqs.~(\ref{exptauedge}),~(\ref{tau2log2}) describing the edge switching.
On the other hand, the prefactors for the edge and corner switching are
qualitatively different, since the latter does not depend on the perimeter
${\mathcal P}$.  This is due to the fact that at $\theta < \pi$ there is
no zero mode, i.e., all ${\tilde \lambda}_{nm}$ except ${\tilde
  \lambda}_{00}$ are positive.  At $\theta\to\pi$ the eigenvalue ${\tilde
  \lambda}_{01}\to 0$, which corresponds to the appearance of a zero mode.
In this case one needs to apply the same procedure as in
Sec.~\ref{large2}, which will lead to the result~(\ref{tau2log2}) for the
prefactor.

Performing the same renormalization~(\ref{newdeltaV}) of $V_{th}$ as in
Secs.~\ref{renormalization2} and~\ref{large2}, we find the expression for
the mean switching time at a corner of angle $\theta$
\begin{equation}
\label{tau_c} 
\tau_c \sim \frac{1}{b\sqrt{\gamma\alpha_R}} 
\exp\left[\frac{4\theta\zeta\eta\alpha_R}{\pi\gamma} \right].
\end{equation}
Note that at $\theta <\pi$ the exponent of $\tau$ for the corner switching
is smaller than that for both interior and edge switching.  This makes
corner switching more efficient far from the threshold.

\section{\label{sec:discussion} Discussion}

In preceding sections we studied the mean time $\tau$ of switching from
the metastable to the stable current state in double-barrier
resonant-tunneling structures.  We calculated both the exponentials and
prefactors of $\tau$ for switching in the small sample regime
[Eq.~(\ref{Tau0})] and for the interior, edge, and corner switching in the
large sample regime [Eqs.~(\ref{tau_i}),~(\ref{tau_e}), and~(\ref{tau_c}),
respectively].  In this section we discuss the dependence of the mean
switching time on voltage for different structural parameters of DBRTS.

We concentrate on the case of round samples, such as the ones used in the
recent experiments.\cite{Grahn98, Teitsworth} As we have shown, when the
voltage $V$ is tuned close to the threshold, the size of critical nucleus
$r_0$ is large compared to the radius of the sample $L$, and the device is
in the small sample regime.  If the voltage is far from $V_{th}$, the
device is in the large sample regime, $L \gg r_0$.  In a typical
experiment $\tau$ is measured in a single device for different values of
bias.  We will therefore assume that all structural parameters and the
size of the sample are fixed, and discuss the switching time as a function
of voltage.  For comparison with experiment we will not distinguish
between $r_R$ and $r_0$ in this section, since the logarithmic in voltage
corrections due to the renormalization of the threshold voltage $V_{th}$
are more challenging to observe.

Our approach is valid as long as the exponents in the expressions for the
switching time, Eqs.~(\ref{Tau0}), (\ref{exptaularge}), and
(\ref{exptauedge}) are much greater than unity.  To check when these
conditions are satisfied, it is convenient to write the exponent in
Eq.~(\ref{exptaularge}) as
\begin{equation}
\label{exp_large}
\frac{8\zeta\eta\alpha}{\gamma}
= \left(\frac{d}{r_0}\right)^{4}.
\end{equation}
Here we introduced a new characteristic length scale
\begin{equation}
\label{d}
d \equiv \left( \frac{8\zeta\eta^{3}}{\gamma^{2}} \right)^{1/4}
\end{equation}
and applied the definition of $r_0$ given by Eq.~(\ref{r0}).  Note that
the length scale $d$ depends on structural parameters of the device, but
not on the sample size or bias.

\begin{figure}
  \resizebox{.46\textwidth}{!}{\includegraphics{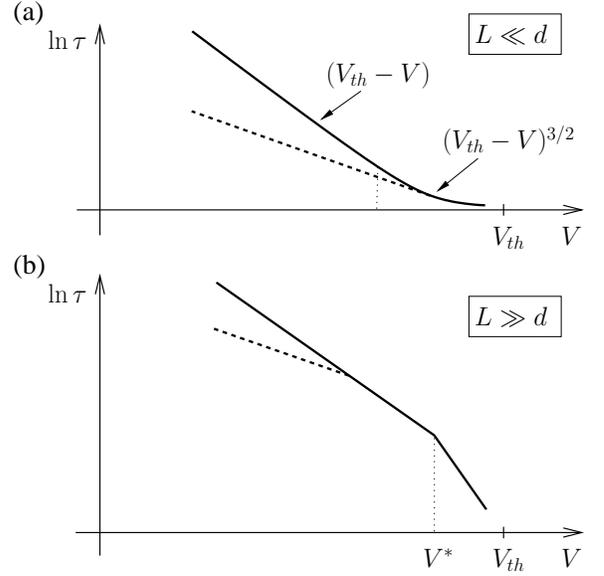}}
\caption{\label{fig:regions} Schematic dependence of the logarithm of the
  mean switching time $\tau$ on voltage.  (a) In round samples at $L\ll
  d$, close to the threshold there is a region of $3/2$-power law
  dependence of $\ln\tau$, followed by the region of linear dependence
  corresponding to the switching at the edge in the regime of large
  sample.  In samples with pronounced corners the region of linear
  dependence corresponds to the switching at the sharpest corner (dashed
  line).  (b) At $L\gg d$, two regions of different linear behavior
  corresponding to the switching in the interior and at the edge of large
  round sample are present.  In samples with pronounced corners these two
  regions are followed by an additional region of linear voltage
  dependence corresponding to the switching at the corner of smallest
  angle $\theta$ shown by dashed line.  The slope of this linear
  dependence is smaller by a factor of $\pi /\theta$ than that of edge
  switching.  }
\end{figure}

Similarly, the exponent of the switching time~(\ref{Tau0}) in a small
sample can be expressed in terms of $d$ and $r_0$ as
\begin{equation}
\label{exp_small}
\frac{4S\alpha^{3/2}}{3\gamma^{1/2}}
= \frac{\pi}{6\zeta} \frac{L^2}{r_{0}^{2}}\left( \frac{d}{r_0}\right)^{4},
\end{equation}
where we used the fact that in round samples $S= \pi L^2$.  This exponent
is much greater than unity at $r_0 \ll (Ld^{2})^{1/3}$.  On the other
hand, the regime of small sample is defined by the condition $r_0\gg L$.
Therefore, it exists only in sufficiently small samples, $L\ll d$.  In
this case close to the threshold there is a region of $3/2$-power law
behavior~(\ref{Tau0}).  As voltage tuned further away from $V_{th}$ (i.e.,
at $r_0 \ll L$), it crosses over to the region of linear voltage
dependence of $\ln\tau$ for the regime of large sample, see solid line in
Fig.~\ref{fig:regions}a.

In large round samples the mean switching time $\tau$ is given by
$\tau^{-1} = \tau_i^{-1} + \tau_e^{-1}$.  Therefore, to find the slope of
linear segment of the curve in Fig.~\ref{fig:regions}a, one has to compare
the rates of switching in the interior and at the edge.  Using
Eqs.~(\ref{tau_i}) and (\ref{tau_e}), the ratio of the rates can be
expressed as
\begin{equation}
\label{Re/Ri}
\frac{\tau_{e}^{-1}}{\tau_{i}^{-1}} \sim
\frac{d}{L}\,
\exp\left[ \frac{1}{2}\left(\frac{d}{r_{0}}\right)^{4}
-3\ln\frac{d}{r_0} \right].
\end{equation}
At $L\ll d$ this result shows that the switching always occurs at the edge
rather than in the interior of the sample.

To summarize, we found that in samples of radius $L \ll d$ starting at
voltage difference $(V_{th}-V)$ corresponding to $r_0 \sim
(Ld^{2})^{1/3}$, one first observes the region of $3/2$-power law
dependence~(\ref{Tau0}) of $\ln\tau$.  Then, as $(V_{th} - V)$ increases,
follows the region of linear dependence~(\ref{tau_e}) corresponding to the
switching at the edge in the regime of large sample, see solid line in
Fig.~\ref{fig:regions}a.

At $L\gg d$ the system is never in the small sample regime.  In this case
the dependence of $\ln\tau$ on voltage is linear, but it may be due to
either interior or edge switching.  According to Eq.~(\ref{Re/Ri}), at
$r_0 \lesssim d$ and very large $L$ interior switching dominates.  At very
small $r_0$ the exponential in Eq.~(\ref{Re/Ri}) becomes very large, and
therefore the switching takes place at the edge.  The crossover voltage
$V^{*}$ between these two regions of linear dependence can be determined
from the condition $\tau_e^{-1} = \tau_i^{-1}$ applied to
Eq.~(\ref{Re/Ri}),
\begin{equation}
\label{V*}
V^{*} \approx V_{th} - \frac{2\eta^2}{\gamma a d^4} \ln\frac{L}{d}.
\end{equation}
Thus, in these large samples the interior switching~(\ref{tau_i})
dominates between $V^{*}$ and $V_{th}$, whereas at voltage below $V^{*}$
the edge switching~(\ref{tau_e}) prevails.  The dependence of $\ln\tau$ on
voltage for $L\gg d$ is shown schematically in Fig.~\ref{fig:regions}b by
solid line.

If the sample size is of order $d$, the dependence of $\ln\tau$ on voltage
can be obtained from the dependences shown in Fig.~\ref{fig:regions}a and
Fig.~\ref{fig:regions}b.  At $L\sim d$ the region of $3/2$-power law
depedence in Fig.~\ref{fig:regions}a and the interior switching region in
Fig.~\ref{fig:regions}b disappear.  Thus, at $L\sim d$ one can only
observe the region of linear voltage dependence corresponding to the edge
switching.

In samples with pronounced corners the dependence of $\ln\tau$ on voltage
is different due to the possibility of corner switching.  The mean
switching time $\tau$ in these samples is given by $\tau^{-1} =
\tau_i^{-1} + \tau_e^{-1} + \tau_c^{-1}$.  At $L\ll d$ an analysis similar
to that for round samples shows that the region of linear dependence
corresponds to the switching at the sharpest corner, Eq.~(\ref{tau_c}).
This dependence is illustrated by dashed line in Fig.~\ref{fig:regions}a.
At $L \gg d$, the two regions of interior and edge switching are followed
by an additional region corresponding to the switching at the sharpest
corner as $(V_{th}-V)$ becomes large, see dashed line in
Fig.~\ref{fig:regions}b.

Depending on the ratio of $L$ and $d$ two qualitatively different voltage
dependences of $\ln\tau$ are expected.  To see whether it is possible to
observe them experimentally, we make a crude estimate of the parameter
$d$.  Substituting $\eta= \sigma/{\tilde c}b$ into Eq.~(\ref{d}) and using
the estimates of $\gamma$ and $B$ found in Appendix~\ref{appendix:AB}, we
get
\begin{equation}
\label{destimate}
d \sim \frac{1}{\sqrt{n}}
\left( \frac{\hbar\sigma}{e^2 T_R} \right)^{3/4} 
\left( \frac{T_L}{T_R} \right)^{1/2}.
\end{equation}
To obtain this expression the capacitance of the device per unit area was
estimated as $c\sim e^2 n/E_F$, and the energy of the level in the well
$E_0$ was assumed to be of the order of $E_F$.  The electron density $n$
in the well is typically of the order of $2\cdot 10^{11}$ $\rm{cm}^{-2}$.
The transmission coefficients $T_{L,R}$ of the left and right barriers can
be varied in the range from 1 to $10^{-4}$, whereas the conductivity
measured in units of $e^2 /\hbar$ varies from 1 to 100.  Assuming $T_L
\sim T_R$, the low bound $d\sim 20$ nm is achieved at $\sigma \sim e^2
/\hbar$ and $T_{L,R}\sim 1$.  The upper bound $d\sim 600$ $\mu$m is
achieved by substituting the maximum value of the conductivity and the
minimum value of the transmission coefficient.  These estimates show that
both the cases of $L\ll d$ and $L\gg d$ are experimentally achievable in
modern DBRTS, as the sample sizes range from 1 to $10^3 \mu$m.

The available experimental data\cite{Teitsworth} confirm that the
dependence of the mean switching time on voltage is indeed exponential.
Based on Eq.~(\ref{destimate}) we estimate $d\sim 10$ $\mu$m, which is
somewhat smaller than the radius of the sample $L= 60$ $\mu$m.  Thus, one
should expect the logarithm of the mean switching time to behave as shown
in Fig.~\ref{fig:regions}b.  (The switching time $\tau$ is referred to as
the \textit{relocation} time in Ref.~\onlinecite{Teitsworth}.)  On the
other hand, it was observed in Ref.~\onlinecite{Teitsworth} that $\ln\tau$
bends upwards, which suggests that $L\ll d$, see Fig.~\ref{fig:regions}a.
One of the possible explanations can be that this experiment was performed
in superlattices, rather than in DBRTS studied in this paper, which makes
our estimate of $d$ unreliable.  To test our theory in more detail, it
would be interesting to carry out similar measurements of $\tau$ in
several samples of different size but with the same structural parameters.
This will ensure that both dependences depicted schematically in
Fig.~\ref{fig:regions}a ($L\ll d$) and Fig.~\ref{fig:regions}b ($L\gg d$)
will be observed.  In addition, the exponential dependence in
Ref.~\onlinecite{Teitsworth} is not very pronounced, since $\tau$ varies
by only one order of magnitude.  This suggests that $\tau$ was measured
rather close to the threshold, and therefore the data captures only the
initial part of either linear dependence for interior switching
[Fig.~\ref{fig:regions}b] or 3/2-power law dependence,
Fig.~\ref{fig:regions}a.  To observe the entire bias dependence shown in
Fig.~\ref{fig:regions}a or Fig.~\ref{fig:regions}b, a measurement of
$\tau$ in a wider range of voltage is needed.

\begin{acknowledgments}
  The authors are grateful to A.~V. Andreev, E.~Sch\"oll, S.~W.
  Teitsworth, and M.~B.  Voloshin for valuable discussions.  The authors
  acknowledge the hospitality of Bell Labs where part of this work was
  carried out.  O.A.T. would like to thank Argonne National Laboratory for
  their hospitality.  This work was supported by the U.S. DOE, Office of
  Science, under Contract No. W-31-109-ENG-38, by the Packard Foundation,
  and by NSF Grant DMR-0214149.
\end{acknowledgments}

\appendix

\section{\label{appendix:AB} Calculation of coefficients $A$ and $B$ in Eq.~(\ref{tunnelingFPE})}

In this appendix we find the functions $A(N)$ and $B(N)$ in
Eq.~(\ref{tunnelingFPE}) in the vicinity of the threshold.  We will assume
that the parameter~(\ref{smalllambda}) is small, $\lambda\ll 1$.  It will
be convenient here to consider $A$ and $B$ as functions of the electron
density $n$ rather than $N=nS$.

Let us write the expression for $A = J_L - J_R$ near the threshold.  On
the upper branch of the $I$-$V$ curve, the level in the well lies within
the conduction band in the left lead; from Eq.~(\ref{ephi}) we obtain
$eV/2 + e^{2}N/2C + E_0 >eV$.  On the lower branch, the level is below the
bottom of the conduction band in the left lead $eV/2 +E_0 < eV$, so that
no current can flow through the well and $N=0$.  Therefore, in the
bistable region
\begin{equation}
\label{in}
eV - \frac{e^{2}N}{2C} < \frac{eV}{2} + E_0 < eV.
\end{equation}
At $\lambda \ll 1$, it follows from Eq.~(\ref{N}) that $e^{2}N/2C$ is
small in comparison with $E_F$ and $E_0$.  One can then see from
Eq.~(\ref{in}) that $E_0 \simeq eV/2$, and Eq.~(\ref{ephi}) results in
$e\phi \simeq eV/2$.  Then from Eq.~(\ref{GR}) we find $\Gamma_R =
\sqrt{2}g_R E_0$.  The expression~(\ref{JL}) for the rate $J_L$ can
also be simplified.  For $\lambda\ll 1$ and $g_{L} \sim g_{R}$, to
first order in $\lambda$ the expression in the square brackets of
Eq.~(\ref{JL}) is $(Sm/\pi\hbar^2)E_F$.  Using Eqs.~(\ref{JR}),~(\ref{JL})
and~(\ref{GL}) with all the above simplifications, close to the threshold
$A=J_L -J_R$ can be approximated as
\begin{eqnarray}
\label{AnV}
A(n, V)& = &\frac{g_{L}}{\hbar}
\sqrt{E_0 \left( E_{0} - \frac{eV}{2} + \frac{e^2 n}{2c} \right) }
\frac{Sm}{\pi\hbar^2} E_{F}
\nonumber \\
&& - \frac{\sqrt{2}g_{R} E_0}{\hbar} nS.
\end{eqnarray}

The density $n$ on the metastable and unstable branches of the $I$-$V$
curve is found by solving the equation $A(n,V) =0$, which reduces to the
quadratic equation
\begin{eqnarray}
\label{squareeq}
&&\left[\frac{e^{2}n}{2c}\right]^{2}
- \left(\lambda \frac{g_{L}}{g_{R}} \right)^{2}
\frac{E_{F}^{2}}{2E_0} \left[\frac{e^{2}n}{2c}\right]
\nonumber \\
&&+ \left(\lambda \frac{g_{L}}{g_{R}} \right)^{2}
\frac{E_{F}^{2}}{2E_0} \left(\frac{eV}{2} - E_0 \right) = 0.
\end{eqnarray}
At the threshold the two solutions for $n$ coincide.  This condition
enables us to find the threshold voltage and density
\begin{eqnarray}
\label{eVth}
V_{th}& = &\frac{2E_0}{e} \left[1 + \frac{\lambda^{2}}{8}
\frac{g_{L}^{2}}{g_{R}^{2}} \frac{E_{F}^{2}}{E_{0}^{2}}\right],
\\
\label{n_th}
n_{th}& = &\frac{\lambda^2}{2} \left(\frac{g_L}{g_R}\right)^{2}
\frac{c}{e^2}\frac{E_F^2}{E_0}.
\end{eqnarray}

Using Eqs.~(\ref{Bconst}),~(\ref{n_th}) and the fact that near the threshold
$\Gamma_R = \sqrt{2}g_R E_0$, we find the value of $B(n)$ at the
threshold:
\begin{equation}
\label{Bnth}
B(n_{th}) = \sqrt{2}\lambda^2 \frac{1}{\hbar}
\frac{g_L^2}{g_R} \frac{C}{e^2}E_F^{2}.
\end{equation}

Using Eqs.~(\ref{eVth}),~(\ref{n_th}) we expand $A(n, V)$ given by
Eq.~(\ref{AnV}) in Taylor series near $(n_{th}, V_{th})$ up to the first
non-vanishing terms in $n-n_{th}$ and $V-V_{th}$, respectively.  At the
threshold $A(n)=0$ has only one solution, i.e., the first derivative with
respect to $n$ equals to zero, and therefore we need to expand up to the
second order in $n$.  The result can be presented as
\begin{subequations}
\label{AandB}
\begin{eqnarray}
\label{A}
A(n)& = & -\frac{B(n_{th})}{2}\left[-\alpha+ \gamma (n-n_{th})^2\right],
\\
 \label{a0coefficient}
\alpha & = & \frac{2}{\lambda^2} \left(\frac{g_R}{g_L}\right)^2
\frac{E_0}{E_F^2}e(V_{th}-V),
\\
\label{b0coefficient}
\gamma & = & \frac{2}{\lambda^4} \left(\frac{g_R}{g_L}\right)^4
\left(\frac{e^2}{c}\right)^2 \frac{E_0^2}{E_F^4}.
\end{eqnarray}
\end{subequations}
Since $dU/dN =-2A/B$, the coefficients $\alpha$ and $\gamma$ coincide with
those used in Eq.~(\ref{cubicPotential}).

Assuming rectangular potential profile in the well, parameters
$g_{L,R}$ can be estimated in terms of the transmission coefficients
of the barriers as $g_{L,R} = T_{L,R}/\pi$.

\section{\label{appendix:balance} Derivation of Eq.~(\ref{detailed})
from the detailed balance principle}

Let us consider two very close to each other points $\textbf{r}_1$ and
$\textbf{r}_2$ in the well.  The system is in a local equilibrium, and the
electron distributions are given by Fermi functions.  We assume electrons
to be sufficiently well coupled to the lattice, so that the temperature
$T$ is the same everywhere in the quantum well.  Then the probabilities of
diffusion between these two points are given by
\begin{equation}
\label{W12}
\begin{aligned}
W_{\Delta t}(\mathbf{r}_1 ,\mathbf{r}_2 ;n) = \sum_{ij} W_{ij}
f_i (1-f_j) = \sum_{ij} W_{ij} f_i f_j e^{\frac{\epsilon_j - \mu_2}{T}},\\
W_{\Delta t}(\mathbf{r}_2 ,\mathbf{r}_1 ;n) = \sum_{ij} W_{ji}
f_j (1-f_i) = \sum_{ij} W_{ji} f_i f_j e^{\frac{\epsilon_i - \mu_1}{T}}.
\end{aligned}
\end{equation}
Here $i$ and $j$ label the energy levels at positions $\textbf{r}_1$ and
$\textbf{r}_2$, respectively; $f_{i(j)}$ are the Fermi functions, and
$W_{ij}$ is the probability of transition from occupied level $i$ to
unoccupied level $j$.

In equilibrium the transition rates satisfy the detailed balance
condition:
\begin{equation}
\label{dbprinciple}
W_{ij} e^{-\epsilon_{i}/T} = W_{ji} e^{-\epsilon_{j}/T}.
\end{equation}
Our system is away from equilibrium, since the electrochemical potential
$\mu(\mathbf{r})$ varies with the electron density $n(\mathbf{r})$.
However, expression~(\ref{dbprinciple}) is still applicable for the
relevant electron scattering processes.  For example, in the case of
elastic scattering by impurities $\epsilon_i = \epsilon_j$, and $W_{ij} =
W_{ji}$ due to time reversal symmetry, so that Eq.~(\ref{dbprinciple})
holds.  Furthermore, one can easily check that for electron-phonon
scattering expression~(\ref{dbprinciple}) is also valid, because the
phonons are not sensitive to the change in electrochemical potential.

Strictly speaking in the presence of electron-electron scattering
expression~(\ref{dbprinciple}) is incorrect.  If electron during the
transition from state $i$ to $j$ scatters off an electron at position
$\mathbf{r}'_1$, the latter moves to position $\mathbf{r}'_2$.  Then one
finds an additional factor of $\exp[(\mu'_1 -\mu'_2)/T]$ in the right-hand
side of Eq.~(\ref{dbprinciple}).  However, because the electron-electron
interaction is screened, the distance $\mathbf{r}'_2 - \mathbf{r}'_1$ is
of the order of the screening length in the well.  The change of $\mu$ at
such short distances is small compared to the temperature, and thus
Eq.~(\ref{dbprinciple}) is still approximately correct.

Applying expression~(\ref{dbprinciple}) to Eqs.~(\ref{W12}) we obtain
Eq.~(\ref{detailed}).  Since during a short time interval $\Delta t$ an
electron can only diffuse over a short distance, the above proof is
sufficient for the purposes of Sec.~\ref{diffusion}.

As an additional remark, let us show that the expression~(\ref{detailed})
also holds at larger distances.  We consider the probability density $W_t
(\mathbf{r}_i ,\mathbf{r}_f ;n)$ of diffusion from point $\textbf{r}_i$ to
a relatively distant point $\textbf{r}_f$.  Let us divide the time
interval $t$ into $N$ small intervals $\Delta t = t /N$.  Then $W_t$ can
be represented in terms of $W_{\Delta t}$ in the following way:
\begin{equation*}
W_t (\textbf{r}_i ,\textbf{r}_f ;n)
= \prod\limits_{k=1}^{N} \int\!\!
W_{\Delta t}(\textbf{r}_k ,\textbf{r}_{k+1} ;n)\,
d\textbf{r}_{k+1},
\end{equation*}
where $\textbf{r}_1 = \textbf{r}_i$ and $\textbf{r}_{N+1} = \textbf{r}_f$.
The distances between the points $\mathbf{r}_k$ and $\mathbf{r}_{k+1}$ are
small, so that the expression~(\ref{detailed}) is applicable.  Since at
small distances $\Delta \mu\ll T$, we can expand Eq.~(\ref{detailed}) up
to the linear terms in $\Delta \mu /T$.  Using this expansion we can
rewrite each integrand in the above expression in terms of $W_{\Delta
t}(\textbf{r}_{k+1} ,\textbf{r}_{k} ;n)$.  Then evaluating the product
over $k$ we obtain Eq.~(\ref{detailed}).  This completes the prove.

\section{\label{appendix:derivationG} Calculation of constant $G$ in Eq.~(\ref{Gdeltat})}

In this appendix we find the constant $G$ in Eq.~(\ref{Gdeltat}) for an
arbitrary scattering mechanism.  This is accomplished by expressing $G$ in
terms of conductivity $\sigma$.

If a small electrochemical potential gradient is applied in the
$x$-direction, it gives rise to an electric current
\begin{equation}
\label{Ohm}
J = - L_y \sigma \frac{\partial}{\partial x} \left( \frac{\mu}{e} \right),
\end{equation}
where $L_y$ is the width of the sample.

Let us find the expression for the current along $x$-axis at $x=0$ in
terms of the transition probability density $W_{\Delta t}$.  It is given
by the difference in the number of electrons crossing the line $x=0$ from
left to right and in the opposite direction in unit time, namely,
\begin{eqnarray}
\label{current}
J & = & \frac{e}{\Delta t}
\int_0^{L_y}\!\! dy_1 \int_0^{L_y}\!\! dy_2
\int_{-\infty}^0 \!\! dx_1 \int_0^{\infty}\!\! dx_2
\nonumber \\
&&\times [ W_{\Delta t}({\bf r}_1 ,{\bf r}_2 ;n)
- W_{\Delta t}({\bf r}_2 , {\bf r}_1 ;n) ].
\end{eqnarray}
In equilibrium, i.e. at $\partial\mu /\partial x =0$, the expression for
the difference of probability densities in the second line of
Eq.~(\ref{current}) vanishes.  Away from equilibrium it can found by using
the ``detailed balance'' expression~(\ref{detailed}):
\begin{equation*}
W_{\Delta t}({\bf r}_1 ,{\bf r}_2 ;n)
- W_{\Delta t}({\bf r}_2 , {\bf r}_1 ;n)
\simeq \frac{\mu_1 - \mu_2}{ T}
W_{\Delta t}({\bf r}_1 ,{\bf r}_2 ;n).
\end{equation*}
Expanding $\mu_1 - \mu_2 \simeq (x_1 -x_2 ) \partial\mu /\partial x$, one
can see that the linearized form of Eq.~(\ref{current}) reproduces
Eq.~(\ref{Ohm}) with the conductivity given by
\begin{eqnarray}
\label{sigma}
\sigma & = &\frac{e^2}{L_y T\Delta t}
\int_0^{L_y}\!\! dy_1 \int_0^{L_y}\!\! dy_2
\int_{-\infty}^0 \!\! dx_1 \int_0^{\infty}\!\! dx_2
\nonumber \\
&&\times (x_2 - x_1 ) W_{\Delta t}({\bf r}_1 ,{\bf r}_2 ;n).
\end{eqnarray}

It is important to note that this expression is taken in the limit
$\partial\mu /\partial x = 0$, so that $W_{\Delta t}$ in Eq.~(\ref{sigma})
is an equilibrium quantity.  Therefore $W_{\Delta t}$ depends only on the
distance between ${\bf r}_{1}$ and ${\bf r}_{2}$, i.e., $W_{\Delta t}({\bf
r}_1 ,{\bf r}_2 ;n) = W_{\Delta t}(|{\bf r}_1 - {\bf r}_2 | ;n)$.  Then
substituting new variables $x =x_2 -x_1$ and $u =(x_1 +x_2)/2$ into the
integral in Eq.~(\ref{sigma}), and integrating over $u$, we find
\begin{eqnarray}
\sigma & = &\frac{e^2}{2 L_y T\Delta t}
\int_0^{L_y}\!\! dy_1 \int_0^{L_y}\!\! dy_2 \int_{-\infty}^{\infty} dx
\nonumber \\
&&\times x^2 W_{\Delta t}( \sqrt{x^2 + (y_1 - y_2)^2} ;n).
\nonumber
\end{eqnarray}

Changing the variables to $y =y_1 -y_2$ and $v =(y_1 +y_2)/2$, and using
the fact that $\int\!\! dv = L_y$, we obtain
\begin{equation}
\label{current1}
\sigma = \frac{e^2}{4T\Delta t} \int\!\! d{\bf r}\,
r^2 W_{\Delta t}(|{\bf r}| ;n).
\end{equation}

Finally, comparing Eqs.~(\ref{Gdeltat}) and~(\ref{current1}) we find $G =
4T\sigma / e^2$.

\section{\label{appendix:tempTerm} Stationary solution of Eq.~(\ref{FPEwithAllTerms}) near the threshold}

In this appendix we discuss the stationary solution $P_s \{n\}$ of the
Fokker-Planck equation~(\ref{FPEwithAllTerms}) near the threshold.  At
bias near $V_{th}$ function $b(n)$ can be approximated by a constant
$b=b(n_{th})$.  Then the equation for $P_s \{n\}$ takes the form,
\begin{equation}
\label{P1}
\left[ u'(n) +  \frac{\delta}{\delta n} - 2\eta \nabla^{2}n
- T\frac{2\sigma}{e^2 b} \nabla^{2} \frac{\delta}{\delta n} \right]
P_s \{n\} = 0.
\end{equation}
Here $\eta = \sigma/{\tilde c}b$ and $2a/b = 2A/B = -u'(n)$, c.f.
Eq.~(\ref{potential}).

It is convenient to present $P_s \{n\}$ in terms of a functional $F_s
\{n\}$, such that $P_s \{n\}= \exp(-F_s \{n\})$.  Then Eq.~(\ref{P1}) takes
the following simple form
\begin{equation}
\label{yf}
- \frac{2\sigma T}{e^{2}b} \nabla^{2}y(\textbf{r})
+ y(\textbf{r}) = f(\textbf{r}),
\end{equation}
where we introduced $y(\textbf{r}) = \delta F_s /\delta n$ and
$f(\mathbf{r}) = u'(n) - 2\eta \nabla^{2}n$.

Solution of this equation is given by
\begin{equation}
\label{yIntegral}
y(\mathbf{r}) = \int d\mathbf{r}'\, f(\textbf{r}')
\mathcal{G} (\mathbf{r} -\mathbf{r}'),
\end{equation}
where $\mathcal{G}$ is presented in terms of the modified Bessel function
$K_0$ as
\begin{equation}
\mathcal{G} (\mathbf{r}) = \frac{1}{2\pi r_T^2} K_0 (r/r_{T}),\quad
r_{T} = \sqrt{\frac{2\sigma T}{e^{2}b}}.
\end{equation}

At low $T$ the characteristic size $r_{T}$ of the Green's function
$\mathcal{G}$ is very small, so that $\mathcal{G}$ can be approximated by
a $\delta$-function.  Then Eq.~(\ref{yIntegral}) greatly simplifies,
\begin{equation}
\label{withoutT}
\frac{\delta}{\delta n}F_{s} \{n\}
= u'(n(\mathbf{r})) - 2\eta\nabla^{2}n(\mathbf{r}).
\end{equation}
The solution of this equation reproduces Eq.~(\ref{P0general}).

The exact criterion for validity of Eq.~(\ref{withoutT}) is given by the
condition $r_T \ll r_0$, where $r_0$ is the characteristic size of the
function $-u'(n) + 2\eta\nabla^{2}n$, see Eq.~(\ref{r0}).  After
substitution of the parameters of the problem from Eqs.~(\ref{AandB})
and~(\ref{r0}) this criterion takes the form:
\begin{equation}
\label{criterion}
T\ll \lambda^2 (1 +\lambda) \left(\frac{g_L}{g_R}\right)^3
\frac{E_{F}^3}{E_{0}^{3/2}\sqrt{e(V_{th} - V)}}.
\end{equation}

To estimate the right-hand side of~(\ref{criterion}) we take the
parameters $\lambda\sim 1$, $E_F \sim E_0$, $e(V_{th} - V) \lesssim E_F$
and $g_L \sim g_R$.  Then the criterion~(\ref{criterion}) reduces to $T\ll
E_F$.  Therefore, one can neglect the temperature term in
Eq.~(\ref{FPEwithAllTerms}) unless the structure is strongly asymmetrical,
so that $g_L \ll g_R$.

\section{\label{appendix:zeta} Properties of the saddle point solution
  $z_{s} (\bm{\rho})$}

In this appendix we derive several relations between integrals involving
$z_s (\bm{\rho})$.  Our goal is to express the integrals $\int (\nabla
z_{s})^{2}\, d\bm{\rho}$ and $\int z_{s}\, d\bm{\rho}$ in terms of $\zeta$
defined by Eq.~(\ref{eq:zeta}).

Integrating Eq.~(\ref{kink}) over the infinite plane and using the fact
that $z_s (\bm{\rho})$ decays rapidly at large $\rho$, we find
\begin{equation}
\label{zeta}
\int\!\! z_{s}\, d\bm{\rho} = \int\!\! z_{s}^2\, d\bm{\rho}.
\end{equation}

To express the integral in Eq.~(\ref{czeromode}) in terms of the constant
$\zeta$, we transform it as
\begin{eqnarray}
\label{rhoint1}
\int\!\! (\nabla z_{s})^{2}\, d\bm{\rho}& = &
- \int\!\! z_{s}\nabla^{2} z_{s}\, d\bm{\rho}
\nonumber \\
& = & -\int\!\! z_{s}^{2}\, d \bm{\rho}
+ \int\!\! z_{s}^{3}\, d\bm{\rho},
\end{eqnarray}
where we used Eq.~(\ref{kink}) to obtain the second line
of~(\ref{rhoint1}).

To express one of the integrals in the second line of Eq.~(\ref{rhoint1})
in terms of the other, we take advantage of the azimuthal symmetry of the
saddle point solution.  Multiplying Eq.~(\ref{zradial}) by $\rho^2 z'_s$
and integrating over $\rho$, we find
\begin{equation*}
\label{rhoint}
\left.\frac{\rho^{2}{z'}_{s}^{2}}{2}\right|_{0}^{\infty}
+\int_{0}^{\infty}\rho^{2}\frac{d}{d\rho}\left(\frac{z_{s}^{3}}{3}
-\frac{z_{s}^{2}}{2}\right)\, d\rho =0.
\end{equation*}
The first term in this equation vanishes, since $z_s \propto e^{-\rho}$ at
$\rho\to\infty$.  The second term can be simplified by integration by
parts, resulting in
\begin{equation}
\label{z2z3}
\int\!\! z_{s}^{3}\, d\bm{\rho}
= \frac{3}{2}\int\!\! z_{s}^{2}\, d\bm{\rho}.
\end{equation}

Using Eqs.~(\ref{zeta})--(\ref{z2z3}) and~(\ref{eq:zeta}) we find the
following expressions for the integrals in Eqs.~(\ref{czeromode})
and~(\ref{final}),
\begin{equation}
\label{nablaz2}
\int\!\! (\nabla z_{s})^{2}\, d\bm{\rho} = 2\zeta, \qquad
\int\!\! z_{s}\, d\bm{\rho} = 4\zeta.
\end{equation}

\section {\label{appendix3} Solutions of Eq.~(\ref{2Dequation})}

In this appendix we find the eigenvalues of continuous spectrum of the
Schr\"odinger equation~(\ref{2Dequation}).  We consider a round sample of
dimensionless radius $R = \sqrt{S /\pi r_0^2}$ with the critical
fluctuation situated in the center.  Note that since we are interested in
the case of large samples, the size of the critical fluctuation is small
compared to the sample size, i.e., $R\gg 1$.

The potential $-2z_s (\rho)$ is azimuthally symmetric, so it is convenient
to solve equation~(\ref{2Dequation}) in polar coordinates.  Separating the
variables in as ${\tilde \phi}_{nm} (\bm{\rho}) = Q_{nm}(\rho) \Psi_{m}
(\varphi)$, we can write the equation for the radial part as follows
\begin{equation}
\label{Qm}
\left[ -\frac{d^{2}}{d\rho^{2}} - \frac{1}{\rho}\frac{d}{d\rho}
+ \frac{m^{2}}{\rho^{2}} - 2z_{s}(\rho)\right] Q_{nm}(\rho)
= q_{nm}^2 Q_{nm}(\rho),
\end{equation}
where $q_{nm}^2 \equiv {\tilde \lambda}_{nm} -1$. This equation is subject
to two boundary conditions: $Q_{nm}(\rho)$ is finite at the origin and
$Q'_{nm}(R)=0$.

Let us first consider an infinite sample.  In the absence of the
attractive potential $-2z_s$, the finite at the origin solutions to
Eq.~(\ref{Qm}) are the Bessel functions of the first kind $J_m (q\rho)$.
Their asymptotic behavior at $\rho\to\infty$ is
\begin{equation}
\label{asJm}
J_{m}(q\rho) \simeq \sqrt{\frac{2}{\pi q\rho}}
\cos\left[q\rho - \frac{\pi}{2}\left(m+\frac{1}{2} \right)\right].
\end{equation}
In the presence of the attractive potential the asymptotic form of the
radial part of the eigenfunction modifies as follows
\begin{equation}
\label{asQm}
Q_{m}(\rho) \simeq \sqrt{\frac{2}{\pi q\rho}}\cos\left[q\rho
- \frac{\pi}{2}\left(m + \frac{1}{2} \right) + \delta_{m}(q) \right].
\end{equation}
Here $\delta_m (q)$ is the scattering phase shift due to the
attractive potential.

For our purposes we only need the expression for $\delta_m$ at large
wavevectors $q$.  At $q\gg 1$ the phase shifts $\delta_m \ll 1$, and can
thus be found in Born approximation,
\begin{equation}
\label{phaseshifts}
\delta_{m}(q) = \pi\int_{0}^{\infty}\!\!
z_{s}(\rho)J_m^2 (q\rho) \rho\, d\rho,
\end{equation}
see also Eq.~(14) in Ref.~\onlinecite{MatveevLarkin}.  Note that
$\delta_m$ is indeed small at $q\gg 1$, because $J_m^2 \propto 1/q$.

In a finite sample the wavevectors $q_{nm}$ are quantized.  Using the
asymptotic form~(\ref{asQm}) and the boundary condition $Q'_{nm}(R)=0$, we
find
\begin{equation}
q_{nm} =
\frac{\pi}{R}\left({\tilde n}
- \frac{\delta_m (q_{nm})}{\pi} \right),
\end{equation}
where ${\tilde n}$ is given by $n+1/4$ if $m$ is even, and by $n+3/4$ if
$m$ is odd, with $n$ being a nonnegative integer.  Then the eigenvalues
${\tilde \lambda}_{nm}$ are given by $1 + q_{nm}^2$.  We use this result
in Sec.~\ref{sec:prefactor} to calculate $\prod_n {\tilde \lambda}_{nm}$.
This product is dominated by the factors with large $q_{nm}$.  Therefore
in Eq.~(\ref{tillambdanm}) we approximate ${\tilde n}$ by the radial
quantum number $n$ and the argument of $\delta_m (q_{nm})$ by $\pi n/R$.

In addition, in Sec.~\ref{sec:prefactor} we need an expression for the sum
of the phase shifts~(\ref{phaseshifts}) over the azimuthal quantum numbers
$m$.  In the right-hand side of Eq.~(\ref{phaseshifts}) only the Bessel
functions $J_{m}(q\rho)$ depend on $m$.  Since the sum of
$J_{m}^{2}(q\rho)$ over $m$ equals unity,~\cite{Gradshteyn} we find
\begin{equation}
\label{final}
\sum_{m=-\infty}^{\infty} \delta_m = \pi \int_{0}^{\infty}\!\!
z_{s}(\rho)\rho\, d\rho = 2\zeta.
\end{equation}
We used Eq.~(\ref{nablaz2}) to express the above integral in terms of the
constant $\zeta$.

This result can also be derived by means of the Friedel sum rule which
states that the sum of the phase shifts in the left-hand side of
Eq.~(\ref{final}) is given by $\pi N$, where $N$ is the average number of
levels in the attractive potential $U=-2z_s(\rho)$.  Since the
two-dimensional density of states $\nu_2 = 1/4\pi$, we find
\begin{equation}
N = \frac{1}{4\pi} \int\!\! 2z_{s}(\rho) d\bm{\rho}.
\end{equation}
Combining this expression with the Friedel sum rule we reproduce the
result~(\ref{final}).

\end{document}